\documentclass[
superscriptaddress,
twocolumn,
showpacs,
bibnotes,
amsmath,
amssymb,
aps,
longbibliography,
prl,
floatfix,
]{revtex4-2}
\usepackage{graphicx} 

\title{FeSi Surface Ferromagnetism}
\date{October 2023}
\usepackage{graphicx}
\usepackage{dcolumn}
\usepackage{bm}
\usepackage{cleveref}
\usepackage{silence}
\usepackage{ulem}
\usepackage{color}
\usepackage{soul}
\usepackage{xcolor}

\newcommand{\Lavg}{$\langle L \rangle$}

\begin{document}

\title{Disordered 2D ferromagnetism at the surface of FeSi} 
\author{Keenan E. Avers}
    \email[]{kavers@umd.edu}
    \affiliation{Maryland Quantum Materials Center and Department of Physics, University of Maryland, College Park, Maryland, USA}

\author{Yun Suk Eo}
    \affiliation{Maryland Quantum Materials Center and Department of Physics, University of Maryland, College Park, Maryland, USA}
    \affiliation{Department of Physics and Astronomy, Texas Tech University, Lubbock, TX 79409, USA}
    
\author{Hyeok Yoon}
    \affiliation{Maryland Quantum Materials Center and Department of Physics, University of Maryland, College Park, Maryland, USA}
    
\author{Jarryd A. Horn}
    \affiliation{Maryland Quantum Materials Center and Department of Physics, University of Maryland, College Park, Maryland, USA}
    
\author{Shanta R. Saha}
    \affiliation{Maryland Quantum Materials Center and Department of Physics, University of Maryland, College Park, Maryland, USA}
    
\author{Alonso Suarez}
    \affiliation{Maryland Quantum Materials Center and Department of Physics, University of Maryland, College Park, Maryland, USA}

\author{Peter Zavalij}
\affiliation{Department of Chemistry, University of Maryland, College Park, Maryland 20742, USA}

\author{Johnpierre Paglione}
    \affiliation{Maryland Quantum Materials Center and Department of Physics, University of Maryland, College Park, Maryland, USA}
    \affiliation{Canadian Institute for Advanced Research, Toronto, Ontario M5G 1Z8, Canada}
    \email{paglione@umd.edu}
    
\date{\today}
    
\begin{abstract}
FeSi is a curious example of a $d$-electron system that manifests many of the same phenomena associated with $f$-electron Kondo insulators, including conducting surface states with potentially non-trivial topology. Here we investigate the magnetization and magnetotransport of these surface states and how a 2D ferromagnetic state at the surface of FeSi influences the surface conductivity. We confirm the 2D ferromagnetism via a systematic study of magnetization 
on groups of filtered fragments with increasing surface area-to-volume ratios, identifying characteristic temperatures and magnetic fields associated with the ordered state. 
The paramagnetic to ferromagnetic transition appears broadened, suggesting disorder, which allows spin fluctuations to manifest up to at least 9 T at 2 K. 
This highlights the need to understand the relation between the disorder of the 2D ferromagnetism and the surface conductivity in FeSi.

\end{abstract}

\maketitle

 
\textit{Introduction}
The notion that metals and insulators are opposite and distinct electronic ground states of materials is so intuitive that even those with non-technical background can appreciate the dichotomy. This familiar concept loses its stable footing in the face of topological insulators \cite{TopInsulatorReviewRevModPhys.83.1057_Qi} in which the bulk is a fully gapped insulator, but the unique character of the band structure gives rise to metallic surface states. The introduction of strong electron correlation between localized $f$-moments and bulk conduction electrons via Kondo hybridization can result in systems known as topological Kondo insulators (TKI) \cite{Dzero2016ARCMPTopoKondInsu} in which even the bulk ground state changes from a high temperature metal to a low temperature insulator with the opening of an energy gap at the Fermi energy.

FeSi is an example of such a system in which strong electron correlations change the ground state, although its classification as a TKI has been controversial \cite{IsFeSiKondo_SCHLESINGER, Mazurenko2010PRBFeSiWithCoTheory, Tomczak_2018} as it is not clear if the 3$d$-electrons of Fe can serve the analogous role of $f$-electrons in rare earth Kondo materials \cite{FZFluxSmB6Inverted, Kushwaha2019Ce343KondoIns, tan2015unconventional, hartstein2018fermi, Liu2022npjQMYbB12QuantumOscilation}. It is a non-magnetic bulk insulator when cooled below room temperature \cite{Arita2008PRBFeSiARPES}, crosses over to a magnetic bulk metal with unusual phonon softening when heated above it \cite{Kahn2022PRBFeSiINSPhononSoftening, Jaccarino1967PhysRev.160.476EarlyFeSiMagnetism}, but can also exhibit evidence of metallic behavior again at very low temperatures \cite{Hunt1994PRBFeSiMagThermTransSbFlux, fang2018evidence}. Recent scanning tunneling microscopy (STM) experiments \cite{Yang_PNAS_2021STS} have investigated the nature of the conducting states and found striking resemblances to the TKI material SmB$_6$ \cite{Zhang2012PhysRevX.3.011011SmB6, Zhu2013PhysRevLett.111.216402SmB6danglingBonds} which hosts conducting surface states.
We recently confirmed that the electrical transport is confined to the surface below $\sim$ 10 K \cite{Eo2023APL_FeSiFeSb2}, but it is unclear if it is a metallic 2D electron gas surface state \cite{fang2018evidence} or an insulating variable range hopping one \cite{Eo2023APL_FeSiFeSb2} depending on whether the sheet conductivity is more or less than the 2D Mott-Ioffe-Regal limit of $\sim$ $e^2/h$ \cite{MIT_DasSarma}. 
It is also interesting to investigate how disorder drives such a transition \cite{Zabrodskii2001PhilMagBColumbgap, SARMA2005SSComm2DMITransition, Huang_2022_2DTopInsMetalTrans, Huang2022PhysRevB.105.054206CoulombDisorder2Dsemiconductors}. In fact, the low temperature electrical resistivity of FeSi appears to vary substantially across its long publication history \cite{FeSi_Schlesinger, FeSiOldElecAndThermCond_BUSCHINGER, SbFluxFeSi_Degiorgi, FeSiTransAnom_Glushkov, FeSiAndersonLocalized_Lisunov, Yang_PNAS_2021STS, MIHALIK_1996_PointContact, Paschen_1997_DilFridge, Changdar_2020_PRBARPESchiral, LUNKENHEIMER_1995_Hopping, DEGIORGI_1995_electrodynamic, OUYANG_2017_CoSub, fang2018evidence, Delaire2015PhysRevB.91.094307IrOsDopping} both qualitatively in apparent functional form and quantitatively by a few orders of magnitude. With confirmation that the low temperature transport is surface dominated, it motivates an investigation of the magnetic properties of the surface.  There have been reports of ferromagnetism in nanowires \cite{Ruiz_2019NanotechFeSiNanowires}, but it was not clear if the magnetic order was 2D until recent thin film work \cite{ThinFeSiZak_Ohtsuka}. It is also unclear how a 2D ferromagnetic order interacts with the conductive surface states and how disorder may play a role between the two, although interest in such reduced dimensional scenarios remains \cite{Burmistrov2018PRB_quantCorr2DMagImp, Mitra2007PRLWeakLocalFeFilm, Lin1996PRB_MR_and_AHE_FeSiLayer}. True 2D ferromagnetic (FM) order is comparatively rare in isotropic systems and is not expected to be stable in the presence of thermal fluctuations owing to the Mermin-Wagner theorem \cite{Mermin1966PhysRevLett.17.1133OGMWtheroem}. The recent discovery of field-controllable 2D van der Waals magnetism in monolayer Cr$_2$Ge$_2$Te$_6$ is a notable exception \cite{gong2017CrGeTevDW2Dferromagnetism}, but the phenomenon remains rare outside of interface or proximity engineering approaches. 

In this work we present magnetotransport results that display evidence of a 2D FM state at the surface of FeSi, despite the low-temperature bulk being non-magnetic aside from impurities \cite{Koyama2000FeSiImpurityMag}. We confirm the existence of the 2D ferromagnetism by careful measurements of magnetization as a function of surface-to-volume ratio controlled by fragment sizes, which allows signatures of surface magnetic moments to be observable from an otherwise bulk measurement technique. The 2D ferromagnetism exhibits broad transitions, which can be taken as signatures of disorder or reduced dimensionality induced spin fluctuations, which may be a factor previously unaccounted for in the sheet conductivity of the surface states in FeSi.

\begin{figure}[]
    \centering
    \includegraphics[scale=0.39]{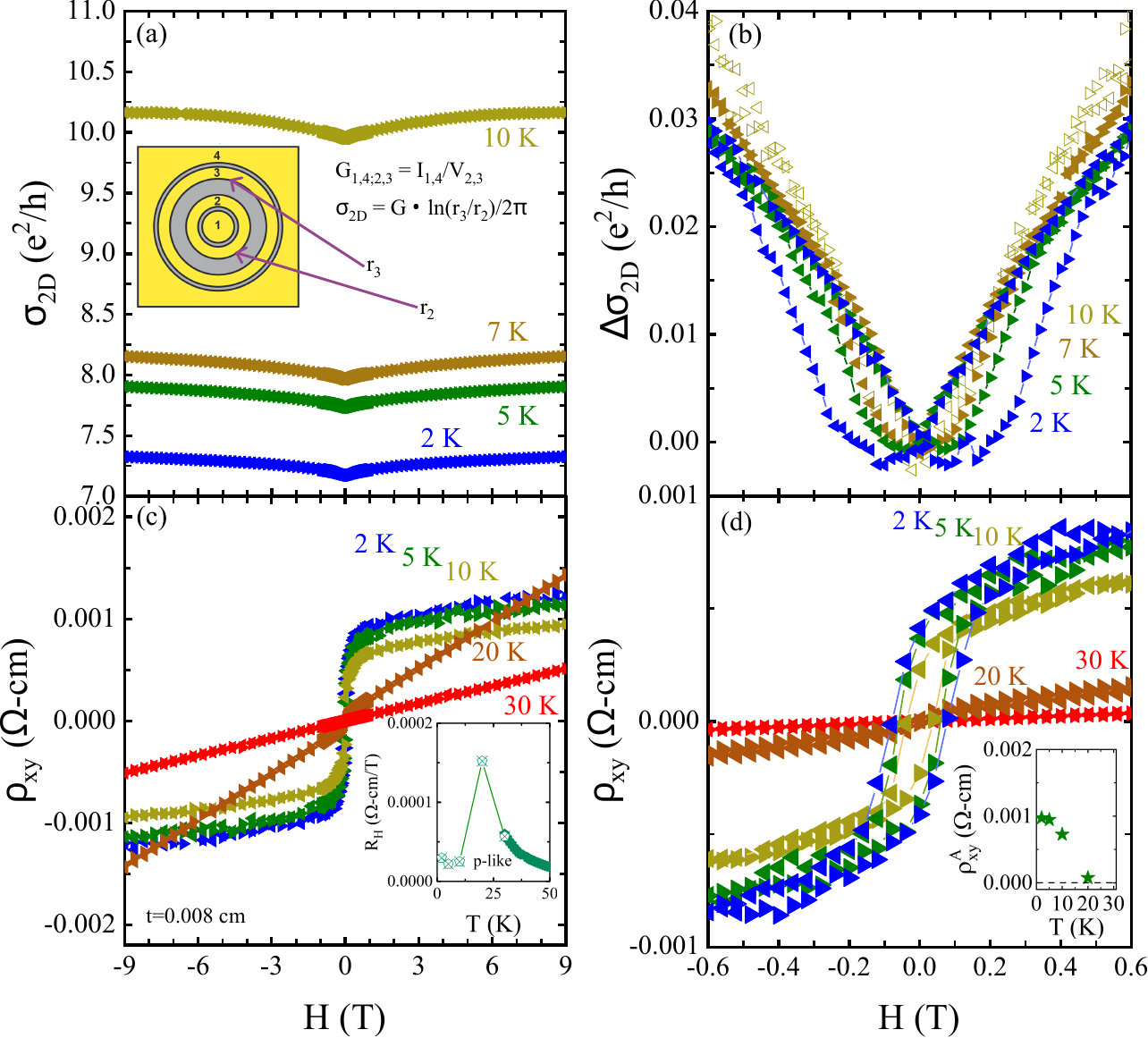}
    \caption{Signatures of FeSi surface ferromagnetism in low temperature 2D conductance ($\sigma_{2D}$) taken via a 4-terminal Corbino geometry on a [111] surface vs. field ($H$) with $H \perp I$ (a). The insets show the contact geometry and associated equations to obtain $\sigma_{2D}$. The change in $\sigma_{2D}$ ($\Delta\sigma_{2D}(H) = \sigma_{2D} (H) -\sigma_{2D} (0)$) (b) showing hysteretic behavior with left-pointing symbols as downsweeps and right-pointing symbols as upsweeps of $H$. Transverse resistivity ($\rho_{xy}$) vs. $H$ exhibiting low-$H$ anomalous and large-$H$ normal Hall effects (c) with normal Hall coefficient ($R_H$) vs. $T$ in the inset. Low-$H$ view of $\rho_{xy}$ showing hysteretic behavior of Hall effect (d) alongside the anomalous Hall resistivity ($\rho^A_{xy}$) vs. $T$ in the inset.}
    \label{fig:Transport}
\end{figure}

FeSi crystals were grown using a variety of methods, including Sn flux \cite{fang2018evidence}, Te flux \cite{Eo2023APL_FeSiFeSb2} and chemical vapor transport with iodine as transport agent \cite{Paschen_1997_DilFridge}. 
Single-crystal X-ray diffraction using Bruker D8Venture w/ PhotonIII diffractometer was performed on each variety, showing that all species result in extremely high quality bulk crystallinity, with final $R_f$ between 1.0 \% to 1.3 \% and lattice constant $a$ between 4.4817 \AA $ $ to 4.4824 \AA $ $ at $T$ = 220 K. 
For this study, Te flux samples were utilized in this study because of their large size and to compare with our previous work \cite{Eo2023APL_FeSiFeSb2}.
Electrical magnetotransport was measured in a commercial cryostat using a 4-terminal Au Corbino disk deposited on the [111] surface, which was polished with 0.3 $\mu$m Al$_2$O$_3$ slurry to prevent subsurface cracks. Magnetization measurements were performed with a commercial SQUID magnetometer using a quartz rod and GE varnish in order to minimize diamagnetic background. Crystals of FeSi were broken apart using blunt force and the resulting fragments filtered using sieves with square meshes of successive opening sizes of 850, 212, 150, 106, 75 and 30 $\mu$m.

\begin{figure}[]
    \centering
    \includegraphics[scale=0.45]{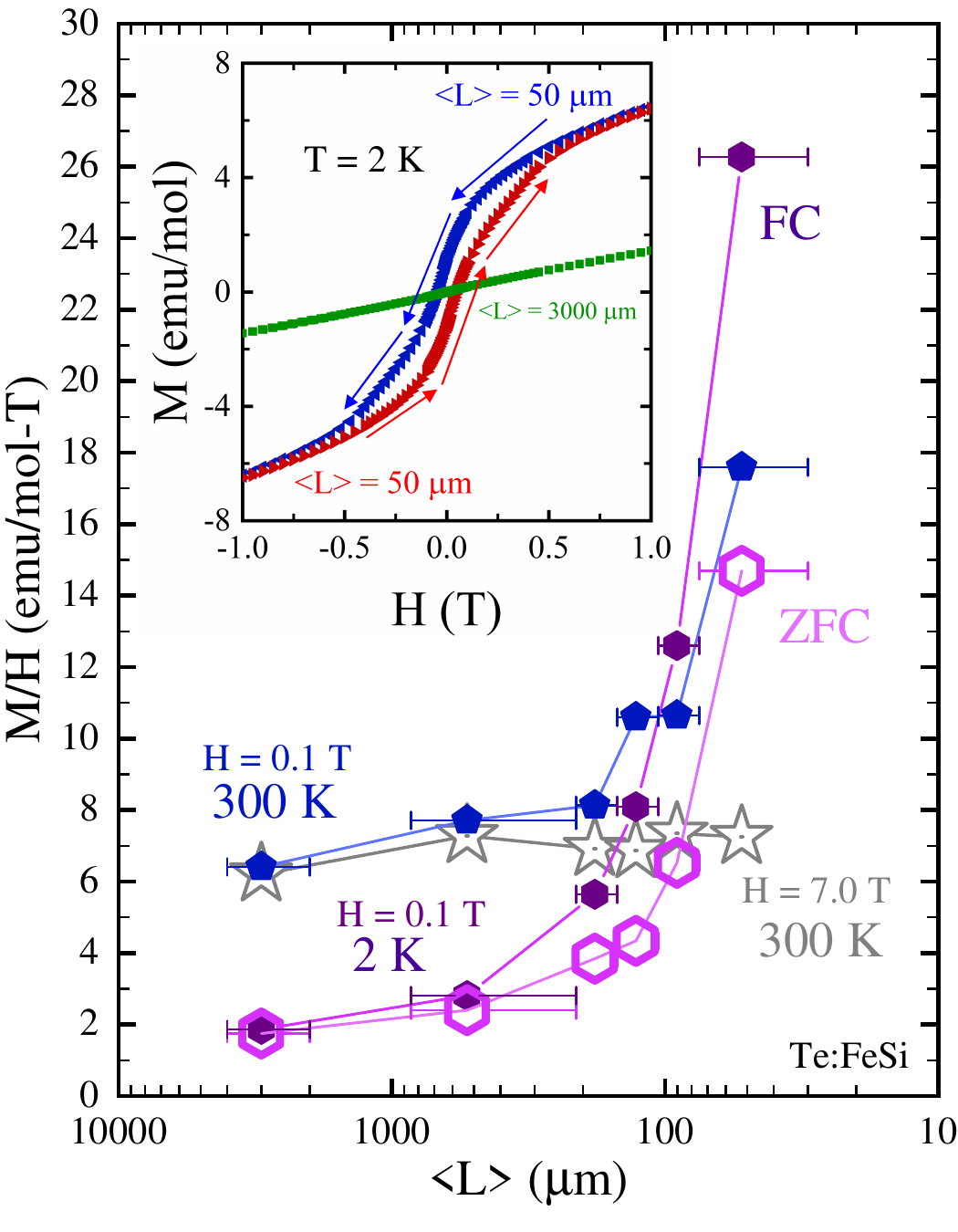}
    \caption{The magnetization ($M$) divided by $H$ ($M/H$) vs. average fragment size (\Lavg) (e). As the total surface area increases, the measured $M/H$ increases at both 300 K and 2 K, which demonstrates the existence of surface magnetic moments distinct from the bulk. The 2 K data exhibits asymmetry between field-cooled (FC) and zero field-cooled (ZFC) measurements and is indicative of a surface ferromagnetic state at low temperatures. The 7.0 T data is independent of fragment size and is mainly from the bulk moments. The inset shows the $M$ vs. $H$ for the intact crystals (\Lavg = 3000 $\mu$m) and the \Lavg = 50 $\mu$m fragments. }
    \label{fig:MagFrag}
\end{figure}

The signatures of 2D ferromagnetism in FeSi are observable via different experimental probes as shown in Fig. \ref{fig:Transport}. The longitudinal transport in Fig. \ref{fig:Transport} (a) shows 2D sheet conductivity ($\sigma_{2D}$) vs. transverse magnetic field ($H$) measured using the Corbino geometry described above.
In the temperature ($T$) range investigated, $\sigma_{2D}$ decreases with decreasing $T$ (ie d$\rho_{2D}$/d$T$ $<$0), although this should not be used to conclude that the surface states are insulating rather than a metallic 2D electron gas \cite{Renard2005PhysRevB.72.075313dirty2DEG, Chen2011NatPhysGrapheneKondo}.  The inset shows the four-terminal Corbino geometry and associated equations to extract $\sigma_{2D}$ from the transport measurements. For all temperatures shown, the 2D magnetoconductivity is positive, which is not expected from the Lorentz force on a 2D electron gas, but is suggestive of a suppression of scattering due to spin fluctuations with increasing field. 
Upon closer examination of the change in $\sigma_{2D}$ ($\Delta\sigma_{2D}(H) = \sigma_{2D} (H) -\sigma_{2D} (0)$) at low $H$ in Fig. \ref{fig:Transport} (b) we observe hysteretic behavior between downsweeps (left-pointing) and upsweeps (right-pointing) of $H$. The hysteretic behavior is also observable in transverse resistivity ($\rho_{xy}$) vs $H$(current along +y, voltage along +x) as a low-$H$ anomalous Hall effect (AHE) \cite{Nagaosa2010RevModPhysAHEreview} in Fig. \ref{fig:Transport} (c) and (d), although we emphasize that quantitative interpretation of $\rho_{xy}$ in the presence of a 2D surface channel and a 3D bulk channel has some subtleties \cite{Kim2013SmB6WedgeHall} that are beyond the scope of this work. 
The zero-field extrapolated AHE ($\rho^A_{xy}$) vs. $T$ and the normal Hall coefficient ($R_H$) vs. $T$ shown in the insets is from linear fitting $\rho_{xy} (H)$ between 4 T and 9 T. The sign of $R_H$ is hole-like (p-like) throughout the entire temperature range.
\begin{figure}[]
    \centering
    \includegraphics[scale=0.3]{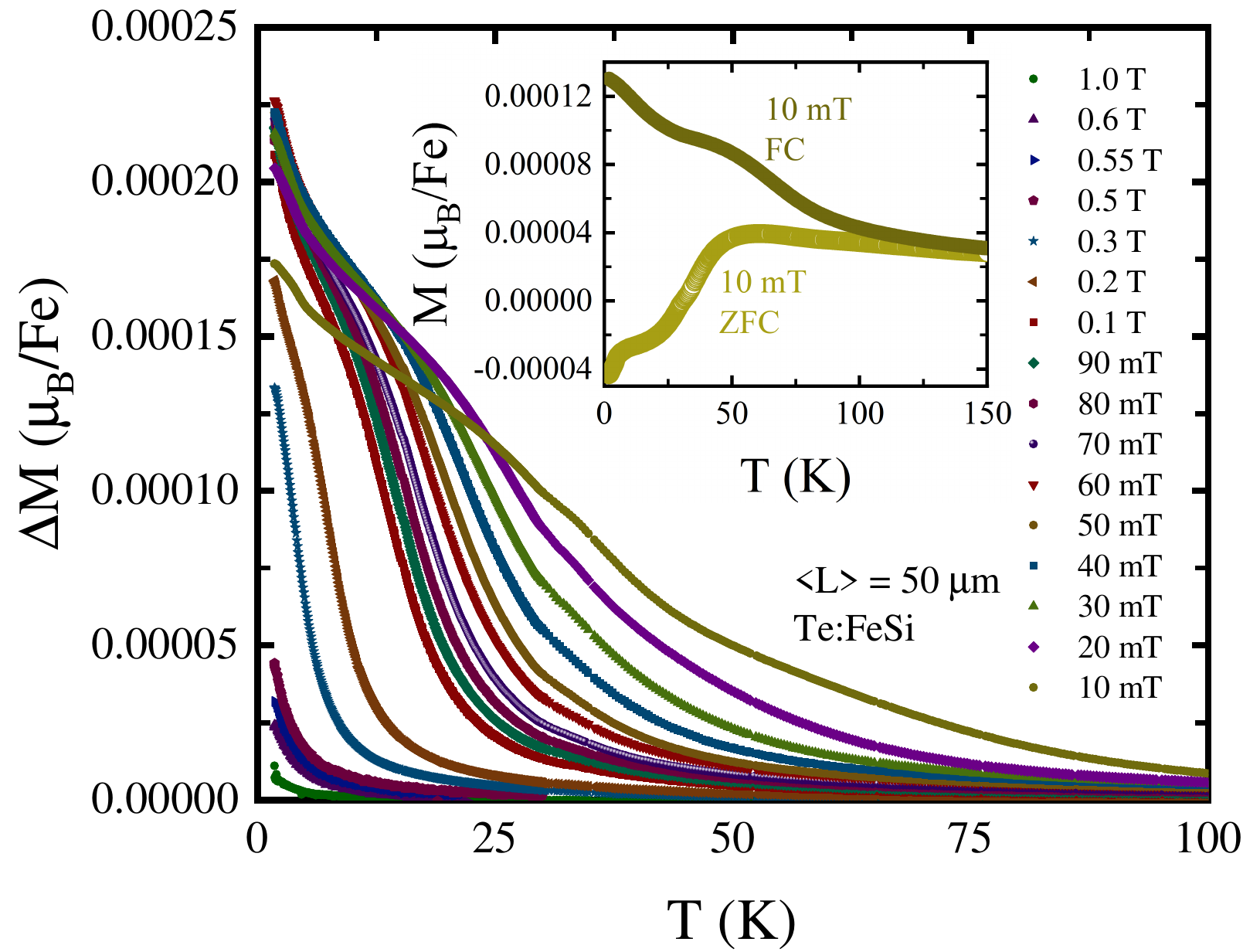}
    \caption{Difference between FC and ZFC magnetization ($\Delta M$) vs. temperature ($T$) of the $<L>$ = 50 $\mu$m sized fragments for several values of $H$. At low magnitudes of $H$ the surface ferromagnetism is prominent as evident by the low $T$ rise of $\Delta M$. At larger $H$ strengths the surface moments are overtaken by the bulk magnetization contribution. The inset shows the FC and ZFC 10 mT data.}
    \label{fig:MvT}
\end{figure}

A relation is apparent between the onset temperature of non-zero $\rho^A_{xy}$ and a maximum of $R_H$ that correlates with the crossover from bulk-dominated to surface-dominated electrical transport that we observed previously \cite{Eo2023APL_FeSiFeSb2} and bears similarity to the same crossover in SmB$_6$ \cite{Kim2013SmB6WedgeHall}. The hysteretic behavior of the surface ferromagnetism observed in Fig. \ref{fig:Transport} (d) is unable to be observed at larger temperatures once the transport is completely shorted by the bulk above 20 K, a limitation that measurements of magnetization ($M$) can overcome. To this end we show the main results of this work in Fig. \ref{fig:MagFrag} as $M/H$ vs average fragment size \Lavg ~filtered through successive sieves. (\Lavg ~is approximated by the midpoint between two sieve mesh openings that define a particular \Lavg ~(ie \Lavg = 50 $\mu m$ are fragments caught by the 30 $\mu m$ sieve but passed through the 75 $ \mu m$ sieve).)  Even at 300 K, the measured moment at 0.1 T systematically increases as \Lavg ~decreases and hence total surface area increases. The 2 K results also not only increase with increasing surface area, but show a widening difference between field-cooled (FC) and zero field-cooled (ZFC) results that is fully consistent with FM order. The 7 T results are nearly independent of \Lavg ~at 300 K, which demonstrates the the bulk magnetic behavior is still present. The inset shows the $M$ vs $H$ at 2 K taken from intact crystals with \Lavg = 3000 $\mu$m as reasonably linear at low $H$, but the upsweeps and downsweeps at the smallest \Lavg = 50 $\mu$m have clear FM hysteretic behavior consistent with the surface transport results and indicative of 2D ferromagnetism at the surface of FeSi.


 The difference between FC and ZFC magnetization ($\Delta M$) vs. $T$ is plotted in Fig. \ref{fig:MvT}, demonstrating a drastic evolution from small to large fields. As the bulk of FeSi is weakly paramagnetic, this subtraction allows any and all bulk contributions to be eliminated. 
 At low fields we observe difference between FC and ZFC results owing to the surface ferromagnetism, but at the largest fields the bulk paramagnetism contributes significantly to the measured moment, as evidenced by the 7 T / 300 K data in Fig. \ref{fig:MagFrag}.
 These results demonstrate that careful measurements of magnetization on a collection of small FeSi crystal fragments at low fields allow investigation of a surface FM state that exists on top of a bulk crystal.

\begin{figure}[]
    \centering
    \includegraphics[scale=0.3]{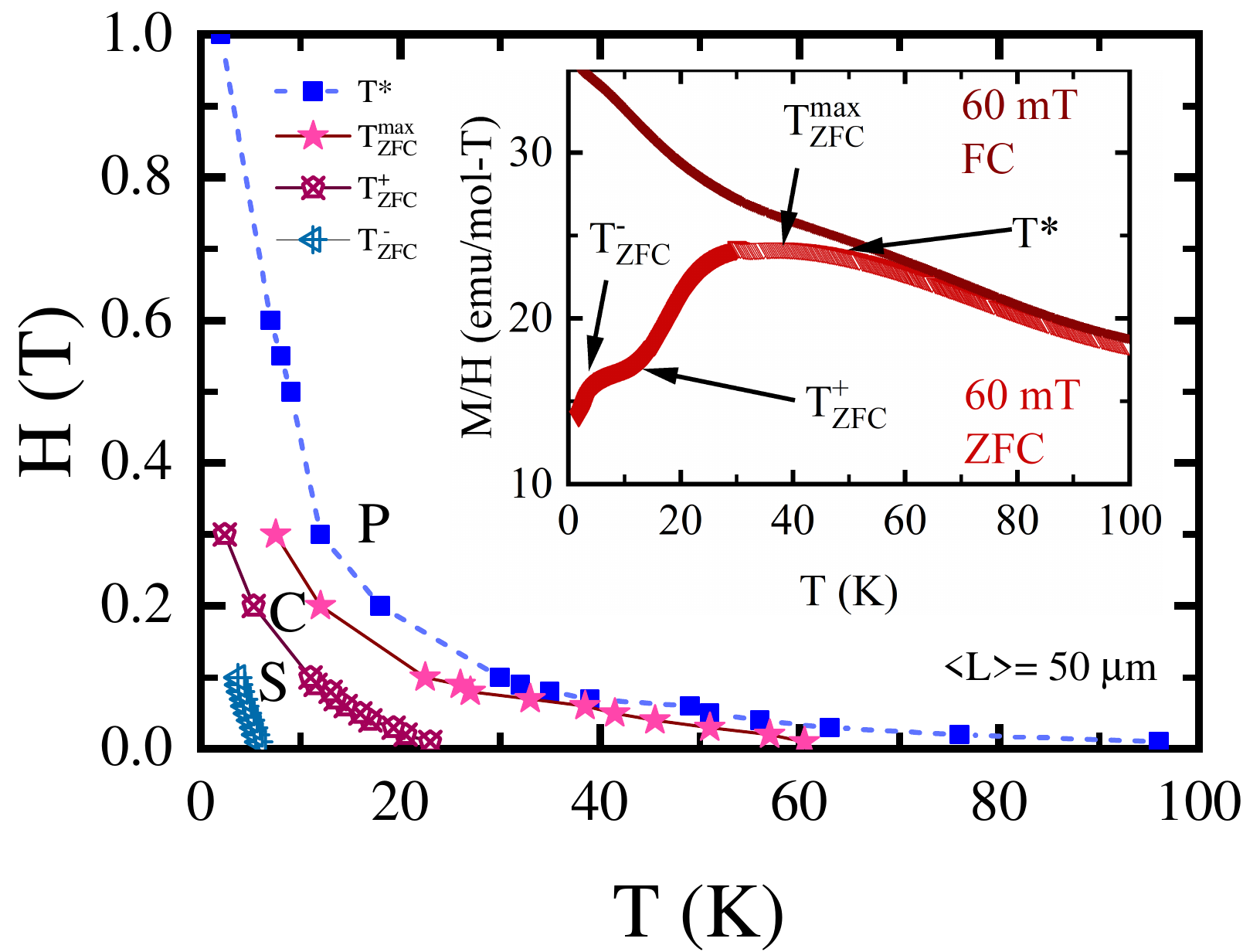}
    \caption{The $H$-$T$ phase diagram for the $<L>$ = 50 $\mu$m sized fragments with the inset showing features used to define tentative transitions from the $M/H$ vs $T$ data. The labeled phases are P (paramagnetic), C (tentative conical state) and S (tentative helical state). The minimal suppression of $T^*$ with $H$ above 0.1 T suggest and that fluctuations persist, likely in a disordered state given the mismatch between $T^{max}_{ZFC}$ and $T^*$.}
    \label{fig:Phase}
\end{figure}

A $T$-$H$ phase diagram based on the \Lavg = 50 $\mu$m fragments is constructed in Fig. \ref{fig:Phase} with the inset showing how the defining features are identified from $M/H$ vs $T$ results. The regions are labeled as P (paramagnetic), C (tentative conical), and S (tentative helical) in keeping with the magnetic phases typical to the 3D magnetically ordered cousins of FeSi \cite{Balas2020PhysRevLett.124.057201CoSiSkyrmions, Munzer2010PhysRevB.81.041203FeCoSiSkyrmion, Bauer2016PhysRevB.93.235144FeCoSiMagHistory}, although the true microscopic nature of these phases will require more investigation.  The presence of a zero field anomalous Hall effect and hysteresis loops in our data are inconsistent with the S phase being a true helical phase as such a state has zero net moment at zero $H$. This is observed experimentally in the anomalous Hall effect of Fe$_{1-x}$Co$_x$Si \cite{Manyala2004NatMatFeSiAHE} that exhibits no hysteresis loops owing to the lack of a zero field spontaneous moment to the helical phase \cite{Beille_1981_JPhysF_FeCoSiHelical, Takeda_2009_JPSJ_FeCoSi_SANS}.  We emphasize that this phase diagram averages over all crystallographic directions, but owing to the cubic symmetry the anisotropy should be minimal and only minor changes in phase boundaries are expected \cite{Bauer2016PhysRevB.93.235144FeCoSiMagHistory}. The signatures associated with the S phase ($T^-_{ZFC}$ and $T^+_{ZFC}$) only appear during a ZFC measurement, while the C phase is the only ordered phase present for a FC measurement, consistent with a report on 3D magnetically ordered Fe$_{1-x}$Co$_x$Si \cite{Bauer2016PhysRevB.93.235144FeCoSiMagHistory} in which the helical phase is a metastable phase. The overall evolution of the phase boundaries with field, as defined by $T^*$ and $T^{max}_{ZFC}$, indicate an unusually slow approach to a field polarized state and suggests that spin fluctuations persist to larger field.

\begin{figure}[]
    \centering
    \includegraphics[scale=0.28]{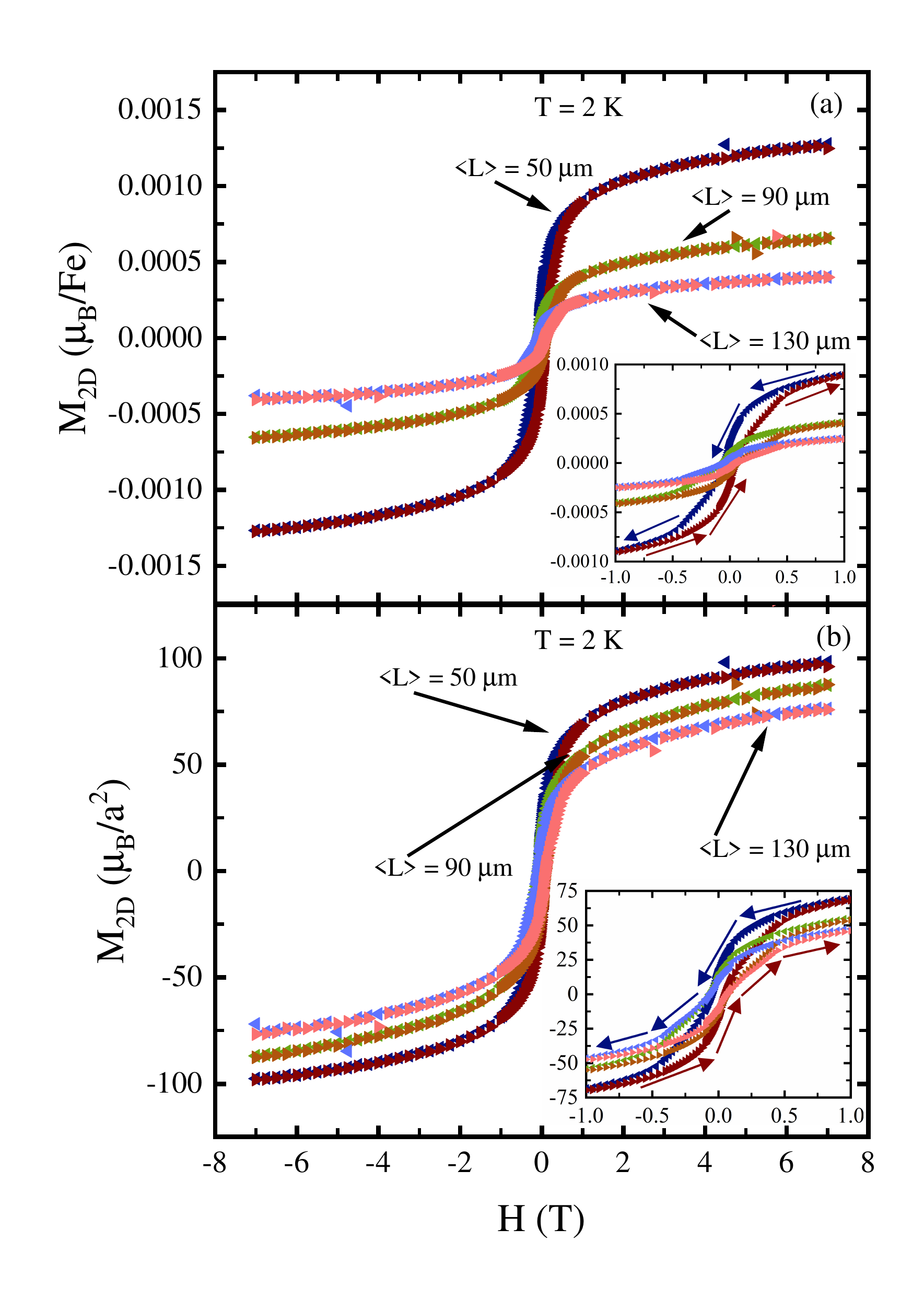}
    \caption{The surface magnetic moment ($M_S$) vs. $H$ at $T$ = 2 K in units of $\mu_B$/Fe (a) or estimated scaling per surface area in units of $\mu_B$/a$^2$ (b) as described in the text with a=4.48 \AA $ $ as the FeSi lattice constant. Hysteretic behavior is observed in the insets with a coercive field on the order of $\sim$ 0.1 T, comparable to the magneto transport results. The surface moments do not completely saturate, even up to 7 T, and suggests remnant ungapped spin fluctuations.}
    \label{fig:MvH}
\end{figure}

The presence of these fluctuations manifest in the 2~K surface magnetization ($M_{2D}$) shown in Fig. \ref{fig:MvH} (a) for a few \Lavg ~values, which is obtained by subtracting the bulk $M(H)$ from the initial intact crystals (\Lavg = 3000 $\mu$m). The hysteretic behavior is resolvable in the low-field data (inset) and is consistent with the surface transport results in Fig. \ref{fig:Transport}. There is a hierarchy with the smaller particle sizes (more surface area) having larger $M_{2D} (H)$, consistent with this being a 2D surface magnetism. In order to scale magnetization from units of $M$/Fe atom to $M$/unit cell surface area with lattice constant $a$ = 4.48 \AA $ $, we postulate that

\begin{equation}
    M_{2D}(\mu_B/a^2)= M_{2D}(\mu_B/Fe)*\left( 4\frac{Fe}{a^3}\right)*
    \frac{\langle L\{a\} \rangle}{6}
    \label{eq:MagAreaScale}
\end{equation}
in which the factor of 4 Fe/$a^3$ converts to magnetization per unit cell volume (4 Fe atoms per unit cell volume $a^3$), and the factor of $\langle L\{a\} \rangle/6$ 
in units of $a$ is the volume-to-surface area ratio of a cube of size \Lavg ~or a sphere of diameter \Lavg. We emphasize that the fragments are not uniform aspect ratio, and in addition Eq. \ref{eq:MagAreaScale} makes no correction for the fraction of fragment surface area normal and perpendicular to applied $H$ nor other demagnetization effects, and hence Eq. \ref{eq:MagAreaScale} has some quantitative inaccuracy. Nevertheless, it is remarkable that the three data sets scale fairly well as shown in Fig. \ref{fig:MvH} (b). As a consistency check, a surface magnetization of 100 $\mu_B/a^2$ $\pm$ 40 $\mu_B/a^2$ alongside an assumed Fe moment of $\sim$ 2.2 $\mu_B$ (dependent of course on exact oxidation state and other microscopic details) implies that the top 11 $\pm$ 5 unit cell layers contribute to the surface moment, a reasonable result given the aforementioned inaccuracies of Eq. \ref{eq:MagAreaScale}, the inherent error on fragment size/geometry, the possibility of minor contribution from naturally occurring Fe oxide \cite{Prakash2007JPhysCondMat}. Recent thin-film results in Ref. \cite{ThinFeSiZak_Ohtsuka} used polarized neutron reflectivity to conclude that the ferromagnetic layer is capable of being only 3 \AA $ $ thick, which is just shy of the FeSi lattice constant, but this thickness, net surface magnetic moment, and easy axis may be tunable in thin films \cite{Hori_2023_AdvMatFeSiCapping, Hori2024PRM_PtCapFeSi} and could naturally be different from the surfaces of our bulk FeSi crystals.

\textit{Discussion} The existence of a 2D FM state at the surface of FeSi has been postulated based on experiments on thin-film FeSi \cite{ThinFeSiZak_Ohtsuka, Manyala2009APLThinFilmFeSi10.1063/1.3152766}, nanowires \cite{Ruiz_2019NanotechFeSiNanowires}, an AHE \cite{Paschen_1997_DilFridge, Manyala2004NatMatFeSiAHE}, and recent microwave signatures \cite{Briendel2023ProbFeSiPublish} alongside angle-resolved MR \cite{Deng2023PhysRevB.108.115158AngleResMR_Sn_FeSi} in Sn flux-grown needles. 
This study, however, is the first conclusive demonstration of surface magnetism residing at the surface of bulk non-magnetic FeSi crystals via direct magnetization. In our case, the low temperature range of the transitions is unusual; for the aforementioned thin films and nanowires the magnetic transitions were observed to occur as high as $\sim$ 200~K without doping, whereas here the transitions are below 100 K, but still higher than $\sim$ 30 K transition in 3D Fe$_{1-x}$Co$_x$Si \cite{Bauer2016PhysRevB.93.235144FeCoSiMagHistory}. 
Disorder is a possible reason for our suppressed transition temperatures in Fig. \ref{fig:Phase} and otherwise broad transitions in Fig. \ref{fig:MvT}.
In particular, the transition temperatures of clean systems almost universally behave in a concave-down manor when suppressed by an external parameter (field, pressure, doping) \cite{Brando2016RevModPhys.88.025006QCP}.  The main phenomena that can break this trend is disorder and the concave-up behavior of our phase boundaries with $H$ suggests that the P-phase in Fig. \ref{fig:Phase} is not a paramagnetic state, but rather exhibits similarities to a Griffiths phase \cite{Saha2022PhysRevB.105.214407GrittithsManganite, DAS2018SolStateComm36SmMnO3DoppedGriffiths}. This is reflected in our $M_S$ vs. $H$ in Fig. \ref{fig:MvH} by a lack of saturating behavior. In fact, 80 K magnetization data (not shown for brevity) is not paramagnetic-like and is qualitatively similar to the 2 K results, albeit with zero coercive field and larger diamagnetic background distortions owing to smaller overall moment, which indicates FM character still persists in our P-phase, even if it lacks long-range order. The source of this disorder is an open question, but possible candidates include standard point defects observable on broken surfaces in STM \cite{Yang_PNAS_2021STS}, although it could be subtle off-stoichiometry of the FeSi crystal \cite{FeSiOldElecAndThermCond_BUSCHINGER, Ohnuma2012ISIJFeSiMetallurgy} which may result in slight cross-occupancy of the Si site by the extra Fe as observed in Co$_{1+x}$Si$_{1-x}$ \cite{Balas2020PhysRevLett.124.057201CoSiSkyrmions}. It is open issue as to why these defects have minimal influence on the bulk behavior, but drastic influence to the surface behavior, but perhaps is a signature of Kondo breakdown on the surface \cite{Alexandrov2015PhysRevLett.114.177202KondoBreakdown}.
The single crystal XRD results could find no evidence for bulk disorder nor significant cross-occupancy between Fe and Si and hence the exact disorder remains a mystery. The alternative is that the spin fluctuations are due entirely to the reduced dimensionality at the surface, but this neglects the history of transport measurements on FeSi. 

How disorder, the 2D magnetism, and surface transport correlate is an open question, especially given the proposed topological nature of surface states in FeSi \cite{Dzero2016ARCMPTopoKondInsu} (although other explanations for the bulk ground state exist \cite{IsFeSiKondo_SCHLESINGER, Mazurenko2010PRBFeSiWithCoTheory, Tomczak_2018}). While surface-based transport in FeSi at low temperatures was only recently confirmed \cite{Eo2023APL_FeSiFeSb2, fang2018evidence}, the long history and variation of transport behavior of FeSi \cite{Eo2023APL_FeSiFeSb2, FeSi_Schlesinger, FeSiOldElecAndThermCond_BUSCHINGER, SbFluxFeSi_Degiorgi, FeSiTransAnom_Glushkov, FeSiAndersonLocalized_Lisunov, Yang_PNAS_2021STS, MIHALIK_1996_PointContact, Paschen_1997_DilFridge, Changdar_2020_PRBARPESchiral, LUNKENHEIMER_1995_Hopping, DEGIORGI_1995_electrodynamic, OUYANG_2017_CoSub, fang2018evidence, Delaire2015PhysRevB.91.094307IrOsDopping} suggests low-temperature transport of FeSi varies excessively without any intuitive explanation. Some of the aforementioned results exhibit saturation of resistivity in the low temperature limit while others continue to become more insulating, depending on the presence/absence of inflection points in $\rho (T)$ and variations in apparent activation energies. Our work suggests that the 2D magnetism at the surface now needs to be considered in any description of the surface transport behavior of FeSi. The notable outlier among FeSi transport results are the Sn flux-grown crystalline needles \cite{fang2018evidence, Briendel2023ProbFeSiPublish, Deng2023PhysRevB.108.115158AngleResMR_Sn_FeSi} in which not only is $\sigma_{2D}$ suggested to be an order of magnitude greater than our Te flux-grown crystals, perhaps $\sim$ 80 $e^2/h$ at 2 K, but also exhibits clean metallic behavior (ie d$\rho_{2D}$/dT $>$0). It is a pressing issue if the 2D magnetism in Sn flux-grown FeSi crystals has significantly less disorder or is otherwise different. Although other TKI systems have signatures of a 2D ferromagnetism at very low temperatures, which was exploited to confirm 1D quantized conductance channels \cite{Nakajima2016NatPhysSmB6Ferromagentism}, it seems unique in the FeSi system as the ferromagnetism occurs at a much higher temperature scale than the onset of the surface states in transport. 
Predictions of FM order in atomically thin layers of the related compound  Fe$_2$Si \cite{Sun2017Fe2SiThinFilm} suggest possible extrinsic origins of the magnetism in FeSi, although no evidence of this phase has been found in the highest quality FeSi surfaces \cite{Yang_PNAS_2021STS}.

\textit{Conclusion}
We have demonstrated the existence of a 2D ferromagnetic state at the surface of bulk crystals of FeSi and its impact on the surface electrical transport. Despite a weak coercive field on the order of $\sim$ 0.1 T at 2 K, the 2D ferromagnetism does not appear to fully polarize and the suppression of spin fluctuations contributes to a positive transverse sheet magnetoconductivity up to at least 9 T. This is supported by our measurements of the surface magnetization, which show no indications of saturation up to 7 T at 2 K. It is an open question as to the topological nature of the surface states, but it still provides reasonable sheet conductivity in excess of 1 $e^2/h$, despite the 2D ferromagnetism exhibiting signatures of disorder, which suggests that the surface state is still a 2D electron gas rather than a localized hopping system. The relation between the disorder of the 2D ferromagnetic state and the surface electrical transport is now a necessity consideration for understanding the $d$-electron topological Kondo insulator candidate FeSi and it drastically varying surface transport results over the decades. 

\begin{acknowledgments}
Research at the University of Maryland was supported by the Gordon and Betty Moore Foundation’s EPiQS
Initiative Grant No. GBMF9071, the Air Force Office of Scientific Research Grant No. FA9950-22-1-0023,
the NIST Center for Neutron Research, and the Maryland Quantum Materials Center. 

\end{acknowledgments} 


\begin{thebibliography}{67}%
\makeatletter
\providecommand \@ifxundefined [1]{%
 \@ifx{#1\undefined}
}%
\providecommand \@ifnum [1]{%
 \ifnum #1\expandafter \@firstoftwo
 \else \expandafter \@secondoftwo
 \fi
}%
\providecommand \@ifx [1]{%
 \ifx #1\expandafter \@firstoftwo
 \else \expandafter \@secondoftwo
 \fi
}%
\providecommand \natexlab [1]{#1}%
\providecommand \enquote  [1]{``#1''}%
\providecommand \bibnamefont  [1]{#1}%
\providecommand \bibfnamefont [1]{#1}%
\providecommand \citenamefont [1]{#1}%
\providecommand \href@noop [0]{\@secondoftwo}%
\providecommand \href [0]{\begingroup \@sanitize@url \@href}%
\providecommand \@href[1]{\@@startlink{#1}\@@href}%
\providecommand \@@href[1]{\endgroup#1\@@endlink}%
\providecommand \@sanitize@url [0]{\catcode `\\12\catcode `\$12\catcode `\&12\catcode `\#12\catcode `\^12\catcode `\_12\catcode `\%12\relax}%
\providecommand \@@startlink[1]{}%
\providecommand \@@endlink[0]{}%
\providecommand \url  [0]{\begingroup\@sanitize@url \@url }%
\providecommand \@url [1]{\endgroup\@href {#1}{\urlprefix }}%
\providecommand \urlprefix  [0]{URL }%
\providecommand \Eprint [0]{\href }%
\providecommand \doibase [0]{https://doi.org/}%
\providecommand \selectlanguage [0]{\@gobble}%
\providecommand \bibinfo  [0]{\@secondoftwo}%
\providecommand \bibfield  [0]{\@secondoftwo}%
\providecommand \translation [1]{[#1]}%
\providecommand \BibitemOpen [0]{}%
\providecommand \bibitemStop [0]{}%
\providecommand \bibitemNoStop [0]{.\EOS\space}%
\providecommand \EOS [0]{\spacefactor3000\relax}%
\providecommand \BibitemShut  [1]{\csname bibitem#1\endcsname}%
\let\auto@bib@innerbib\@empty
\bibitem [{\citenamefont {Qi}\ and\ \citenamefont {Zhang}(2011)}]{TopInsulatorReviewRevModPhys.83.1057_Qi}%
  \BibitemOpen
  \bibfield  {author} {\bibinfo {author} {\bibfnamefont {X.-L.}\ \bibnamefont {Qi}}\ and\ \bibinfo {author} {\bibfnamefont {S.-C.}\ \bibnamefont {Zhang}},\ }\bibfield  {title} {\bibinfo {title} {{Topological insulators and superconductors}},\ }\href {https://doi.org/10.1103/RevModPhys.83.1057} {\bibfield  {journal} {\bibinfo  {journal} {Rev. Mod. Phys.}\ }\textbf {\bibinfo {volume} {83}},\ \bibinfo {pages} {1057} (\bibinfo {year} {2011})}\BibitemShut {NoStop}%
\bibitem [{\citenamefont {Dzero}\ \emph {et~al.}(2016)\citenamefont {Dzero}, \citenamefont {Xia}, \citenamefont {Galitski},\ and\ \citenamefont {Coleman}}]{Dzero2016ARCMPTopoKondInsu}%
  \BibitemOpen
  \bibfield  {author} {\bibinfo {author} {\bibfnamefont {M.}~\bibnamefont {Dzero}}, \bibinfo {author} {\bibfnamefont {J.}~\bibnamefont {Xia}}, \bibinfo {author} {\bibfnamefont {V.}~\bibnamefont {Galitski}},\ and\ \bibinfo {author} {\bibfnamefont {P.}~\bibnamefont {Coleman}},\ }\bibfield  {title} {\bibinfo {title} {Topological kondo insulators},\ }\href {https://doi.org/10.1146/annurev-conmatphys-031214-014749} {\bibfield  {journal} {\bibinfo  {journal} {Annual Review of Condensed Matter Physics}\ }\textbf {\bibinfo {volume} {7}},\ \bibinfo {pages} {249} (\bibinfo {year} {2016})},\ \Eprint {https://arxiv.org/abs/https://doi.org/10.1146/annurev-conmatphys-031214-014749} {https://doi.org/10.1146/annurev-conmatphys-031214-014749} \BibitemShut {NoStop}%
\bibitem [{\citenamefont {Schlesinger}\ \emph {et~al.}(1997)\citenamefont {Schlesinger}, \citenamefont {Fisk}, \citenamefont {Zhang},\ and\ \citenamefont {Maple}}]{IsFeSiKondo_SCHLESINGER}%
  \BibitemOpen
  \bibfield  {author} {\bibinfo {author} {\bibfnamefont {Z.}~\bibnamefont {Schlesinger}}, \bibinfo {author} {\bibfnamefont {Z.}~\bibnamefont {Fisk}}, \bibinfo {author} {\bibfnamefont {H.-T.}\ \bibnamefont {Zhang}},\ and\ \bibinfo {author} {\bibfnamefont {M.}~\bibnamefont {Maple}},\ }\bibfield  {title} {\bibinfo {title} {{Is FeSi a Kondo insulator?}},\ }\href {https://doi.org/https://doi.org/10.1016/S0921-4526(97)00137-3} {\bibfield  {journal} {\bibinfo  {journal} {Physica B: Condensed Matter}\ }\textbf {\bibinfo {volume} {237-238}},\ \bibinfo {pages} {460} (\bibinfo {year} {1997})},\ \bibinfo {note} {proceedings of the Yamada Conference XLV, the International Conference on the Physics of Transition Metals}\BibitemShut {NoStop}%
\bibitem [{\citenamefont {Mazurenko}\ \emph {et~al.}(2010)\citenamefont {Mazurenko}, \citenamefont {Shorikov}, \citenamefont {Lukoyanov}, \citenamefont {Kharlov}, \citenamefont {Gorelov}, \citenamefont {Lichtenstein},\ and\ \citenamefont {Anisimov}}]{Mazurenko2010PRBFeSiWithCoTheory}%
  \BibitemOpen
  \bibfield  {author} {\bibinfo {author} {\bibfnamefont {V.~V.}\ \bibnamefont {Mazurenko}}, \bibinfo {author} {\bibfnamefont {A.~O.}\ \bibnamefont {Shorikov}}, \bibinfo {author} {\bibfnamefont {A.~V.}\ \bibnamefont {Lukoyanov}}, \bibinfo {author} {\bibfnamefont {K.}~\bibnamefont {Kharlov}}, \bibinfo {author} {\bibfnamefont {E.}~\bibnamefont {Gorelov}}, \bibinfo {author} {\bibfnamefont {A.~I.}\ \bibnamefont {Lichtenstein}},\ and\ \bibinfo {author} {\bibfnamefont {V.~I.}\ \bibnamefont {Anisimov}},\ }\bibfield  {title} {\bibinfo {title} {{Metal-insulator transitions and magnetism in correlated band insulators: FeSi and ${\text{Fe}}_{1\ensuremath{-}x}{\text{Co}}_{x}\text{Si}$}},\ }\href {https://doi.org/10.1103/PhysRevB.81.125131} {\bibfield  {journal} {\bibinfo  {journal} {Phys. Rev. B}\ }\textbf {\bibinfo {volume} {81}},\ \bibinfo {pages} {125131} (\bibinfo {year} {2010})}\BibitemShut {NoStop}%
\bibitem [{\citenamefont {Tomczak}(2018)}]{Tomczak_2018}%
  \BibitemOpen
  \bibfield  {author} {\bibinfo {author} {\bibfnamefont {J.~M.}\ \bibnamefont {Tomczak}},\ }\bibfield  {title} {\bibinfo {title} {{Thermoelectricity in correlated narrow-gap semiconductors}},\ }\href {https://doi.org/10.1088/1361-648X/aab284} {\bibfield  {journal} {\bibinfo  {journal} {Journal of Physics: Condensed Matter}\ }\textbf {\bibinfo {volume} {30}},\ \bibinfo {pages} {183001} (\bibinfo {year} {2018})}\BibitemShut {NoStop}%
\bibitem [{\citenamefont {Eo}\ \emph {et~al.}(2021)\citenamefont {Eo}, \citenamefont {Rakoski}, \citenamefont {Sinha}, \citenamefont {Mihaliov}, \citenamefont {Fuhrman}, \citenamefont {Saha}, \citenamefont {Rosa}, \citenamefont {Fisk}, \citenamefont {Hatnean}, \citenamefont {Balakrishnan}, \citenamefont {Chamorro}, \citenamefont {Phelan}, \citenamefont {Koohpayeh}, \citenamefont {McQueen}, \citenamefont {Kang}, \citenamefont {Song}, \citenamefont {Cho}, \citenamefont {Fuhrer}, \citenamefont {Paglione},\ and\ \citenamefont {Kurdak}}]{FZFluxSmB6Inverted}%
  \BibitemOpen
  \bibfield  {author} {\bibinfo {author} {\bibfnamefont {Y.~S.}\ \bibnamefont {Eo}}, \bibinfo {author} {\bibfnamefont {A.}~\bibnamefont {Rakoski}}, \bibinfo {author} {\bibfnamefont {S.}~\bibnamefont {Sinha}}, \bibinfo {author} {\bibfnamefont {D.}~\bibnamefont {Mihaliov}}, \bibinfo {author} {\bibfnamefont {W.~T.}\ \bibnamefont {Fuhrman}}, \bibinfo {author} {\bibfnamefont {S.~R.}\ \bibnamefont {Saha}}, \bibinfo {author} {\bibfnamefont {P.~F.~S.}\ \bibnamefont {Rosa}}, \bibinfo {author} {\bibfnamefont {Z.}~\bibnamefont {Fisk}}, \bibinfo {author} {\bibfnamefont {M.~C.}\ \bibnamefont {Hatnean}}, \bibinfo {author} {\bibfnamefont {G.}~\bibnamefont {Balakrishnan}}, \bibinfo {author} {\bibfnamefont {J.~R.}\ \bibnamefont {Chamorro}}, \bibinfo {author} {\bibfnamefont {W.~A.}\ \bibnamefont {Phelan}}, \bibinfo {author} {\bibfnamefont {S.~M.}\ \bibnamefont {Koohpayeh}}, \bibinfo {author} {\bibfnamefont {T.~M.}\ \bibnamefont {McQueen}}, \bibinfo {author} {\bibfnamefont {B.}~\bibnamefont {Kang}}, \bibinfo {author}
  {\bibfnamefont {M.-s.}\ \bibnamefont {Song}}, \bibinfo {author} {\bibfnamefont {B.}~\bibnamefont {Cho}}, \bibinfo {author} {\bibfnamefont {M.~S.}\ \bibnamefont {Fuhrer}}, \bibinfo {author} {\bibfnamefont {J.}~\bibnamefont {Paglione}},\ and\ \bibinfo {author} {\bibfnamefont {C.}~\bibnamefont {Kurdak}},\ }\bibfield  {title} {\bibinfo {title} {{Bulk transport paths through defects in floating zone and Al flux grown ${\mathrm{SmB}}_{6}$}},\ }\href {https://doi.org/10.1103/PhysRevMaterials.5.055001} {\bibfield  {journal} {\bibinfo  {journal} {Phys. Rev. Materials}\ }\textbf {\bibinfo {volume} {5}},\ \bibinfo {pages} {055001} (\bibinfo {year} {2021})}\BibitemShut {NoStop}%
\bibitem [{\citenamefont {Kushwaha}\ \emph {et~al.}(2019)\citenamefont {Kushwaha}, \citenamefont {Chan}, \citenamefont {Park}, \citenamefont {Thomas}, \citenamefont {Bauer}, \citenamefont {Thompson}, \citenamefont {Ronning}, \citenamefont {Rosa},\ and\ \citenamefont {Harrison}}]{Kushwaha2019Ce343KondoIns}%
  \BibitemOpen
  \bibfield  {author} {\bibinfo {author} {\bibfnamefont {S.~K.}\ \bibnamefont {Kushwaha}}, \bibinfo {author} {\bibfnamefont {M.~K.}\ \bibnamefont {Chan}}, \bibinfo {author} {\bibfnamefont {J.}~\bibnamefont {Park}}, \bibinfo {author} {\bibfnamefont {S.~M.}\ \bibnamefont {Thomas}}, \bibinfo {author} {\bibfnamefont {E.~D.}\ \bibnamefont {Bauer}}, \bibinfo {author} {\bibfnamefont {J.~D.}\ \bibnamefont {Thompson}}, \bibinfo {author} {\bibfnamefont {F.}~\bibnamefont {Ronning}}, \bibinfo {author} {\bibfnamefont {P.~F.~S.}\ \bibnamefont {Rosa}},\ and\ \bibinfo {author} {\bibfnamefont {N.}~\bibnamefont {Harrison}},\ }\bibfield  {title} {\bibinfo {title} {{Magnetic field-tuned Fermi liquid in a Kondo insulator}},\ }\href@noop {} {\bibfield  {journal} {\bibinfo  {journal} {Nature communications}\ }\textbf {\bibinfo {volume} {10}},\ \bibinfo {pages} {5487} (\bibinfo {year} {2019})}\BibitemShut {NoStop}%
\bibitem [{\citenamefont {Tan}\ \emph {et~al.}(2015)\citenamefont {Tan}, \citenamefont {Hsu}, \citenamefont {Zeng}, \citenamefont {Hatnean}, \citenamefont {Harrison}, \citenamefont {Zhu}, \citenamefont {Hartstein}, \citenamefont {Kiourlappou}, \citenamefont {Srivastava}, \citenamefont {Johannes} \emph {et~al.}}]{tan2015unconventional}%
  \BibitemOpen
  \bibfield  {author} {\bibinfo {author} {\bibfnamefont {B.}~\bibnamefont {Tan}}, \bibinfo {author} {\bibfnamefont {Y.-T.}\ \bibnamefont {Hsu}}, \bibinfo {author} {\bibfnamefont {B.}~\bibnamefont {Zeng}}, \bibinfo {author} {\bibfnamefont {M.~C.}\ \bibnamefont {Hatnean}}, \bibinfo {author} {\bibfnamefont {N.}~\bibnamefont {Harrison}}, \bibinfo {author} {\bibfnamefont {Z.}~\bibnamefont {Zhu}}, \bibinfo {author} {\bibfnamefont {M.}~\bibnamefont {Hartstein}}, \bibinfo {author} {\bibfnamefont {M.}~\bibnamefont {Kiourlappou}}, \bibinfo {author} {\bibfnamefont {A.}~\bibnamefont {Srivastava}}, \bibinfo {author} {\bibfnamefont {M.}~\bibnamefont {Johannes}}, \emph {et~al.},\ }\bibfield  {title} {\bibinfo {title} {{Unconventional Fermi surface in an insulating state}},\ }\href@noop {} {\bibfield  {journal} {\bibinfo  {journal} {Science}\ }\textbf {\bibinfo {volume} {349}},\ \bibinfo {pages} {287} (\bibinfo {year} {2015})}\BibitemShut {NoStop}%
\bibitem [{\citenamefont {Hartstein}\ \emph {et~al.}(2018)\citenamefont {Hartstein}, \citenamefont {Toews}, \citenamefont {Hsu}, \citenamefont {Zeng}, \citenamefont {Chen}, \citenamefont {Hatnean}, \citenamefont {Zhang}, \citenamefont {Nakamura}, \citenamefont {Padgett}, \citenamefont {Rodway-Gant} \emph {et~al.}}]{hartstein2018fermi}%
  \BibitemOpen
  \bibfield  {author} {\bibinfo {author} {\bibfnamefont {M.}~\bibnamefont {Hartstein}}, \bibinfo {author} {\bibfnamefont {W.}~\bibnamefont {Toews}}, \bibinfo {author} {\bibfnamefont {Y.-T.}\ \bibnamefont {Hsu}}, \bibinfo {author} {\bibfnamefont {B.}~\bibnamefont {Zeng}}, \bibinfo {author} {\bibfnamefont {X.}~\bibnamefont {Chen}}, \bibinfo {author} {\bibfnamefont {M.~C.}\ \bibnamefont {Hatnean}}, \bibinfo {author} {\bibfnamefont {Q.}~\bibnamefont {Zhang}}, \bibinfo {author} {\bibfnamefont {S.}~\bibnamefont {Nakamura}}, \bibinfo {author} {\bibfnamefont {A.}~\bibnamefont {Padgett}}, \bibinfo {author} {\bibfnamefont {G.}~\bibnamefont {Rodway-Gant}}, \emph {et~al.},\ }\bibfield  {title} {\bibinfo {title} {{Fermi surface in the absence of a Fermi liquid in the Kondo insulator SmB$_6$}},\ }\href@noop {} {\bibfield  {journal} {\bibinfo  {journal} {Nature Physics}\ }\textbf {\bibinfo {volume} {14}},\ \bibinfo {pages} {166} (\bibinfo {year} {2018})}\BibitemShut {NoStop}%
\bibitem [{\citenamefont {Liu}\ \emph {et~al.}(2022)\citenamefont {Liu}, \citenamefont {Hickey}, \citenamefont {Hartstein}, \citenamefont {Davies}, \citenamefont {Eaton}, \citenamefont {Elvin}, \citenamefont {Polyakov}, \citenamefont {Vu}, \citenamefont {Wichitwechkarn}, \citenamefont {Forster}, \citenamefont {Wosnitza}, \citenamefont {Murphy}, \citenamefont {Shitsevalova}, \citenamefont {Johannes}, \citenamefont {Hatnean}, \citenamefont {Balakrishnan}, \citenamefont {Lonzarich},\ and\ \citenamefont {ebastian}}]{Liu2022npjQMYbB12QuantumOscilation}%
  \BibitemOpen
  \bibfield  {author} {\bibinfo {author} {\bibfnamefont {H.}~\bibnamefont {Liu}}, \bibinfo {author} {\bibfnamefont {A.~J.}\ \bibnamefont {Hickey}}, \bibinfo {author} {\bibfnamefont {M.}~\bibnamefont {Hartstein}}, \bibinfo {author} {\bibfnamefont {A.~J.}\ \bibnamefont {Davies}}, \bibinfo {author} {\bibfnamefont {A.~G.}\ \bibnamefont {Eaton}}, \bibinfo {author} {\bibfnamefont {T.}~\bibnamefont {Elvin}}, \bibinfo {author} {\bibfnamefont {E.}~\bibnamefont {Polyakov}}, \bibinfo {author} {\bibfnamefont {T.~H.}\ \bibnamefont {Vu}}, \bibinfo {author} {\bibfnamefont {V.}~\bibnamefont {Wichitwechkarn}}, \bibinfo {author} {\bibfnamefont {T.}~\bibnamefont {Forster}}, \bibinfo {author} {\bibfnamefont {J.}~\bibnamefont {Wosnitza}}, \bibinfo {author} {\bibfnamefont {T.~P.}\ \bibnamefont {Murphy}}, \bibinfo {author} {\bibfnamefont {N.}~\bibnamefont {Shitsevalova}}, \bibinfo {author} {\bibfnamefont {M.~D.}\ \bibnamefont {Johannes}}, \bibinfo {author} {\bibfnamefont {M.~C.}\ \bibnamefont {Hatnean}}, \bibinfo {author}
  {\bibfnamefont {G.}~\bibnamefont {Balakrishnan}}, \bibinfo {author} {\bibfnamefont {G.~G.}\ \bibnamefont {Lonzarich}},\ and\ \bibinfo {author} {\bibfnamefont {S.~E.}\ \bibnamefont {ebastian}},\ }\bibfield  {title} {\bibinfo {title} {{f-electron hybridised Fermi surface in magnetic field-induced metallic YbB$_{12}$}},\ }\href {https://doi.org/10.1038/s41535-021-00413-7} {\bibfield  {journal} {\bibinfo  {journal} {npj Quantum Materials}\ }\textbf {\bibinfo {volume} {7}},\ \bibinfo {pages} {12} (\bibinfo {year} {2022})}\BibitemShut {NoStop}%
\bibitem [{\citenamefont {Arita}\ \emph {et~al.}(2008)\citenamefont {Arita}, \citenamefont {Shimada}, \citenamefont {Takeda}, \citenamefont {Nakatake}, \citenamefont {Namatame}, \citenamefont {Taniguchi}, \citenamefont {Negishi}, \citenamefont {Oguchi}, \citenamefont {Saitoh}, \citenamefont {Fujimori},\ and\ \citenamefont {Kanomata}}]{Arita2008PRBFeSiARPES}%
  \BibitemOpen
  \bibfield  {author} {\bibinfo {author} {\bibfnamefont {M.}~\bibnamefont {Arita}}, \bibinfo {author} {\bibfnamefont {K.}~\bibnamefont {Shimada}}, \bibinfo {author} {\bibfnamefont {Y.}~\bibnamefont {Takeda}}, \bibinfo {author} {\bibfnamefont {M.}~\bibnamefont {Nakatake}}, \bibinfo {author} {\bibfnamefont {H.}~\bibnamefont {Namatame}}, \bibinfo {author} {\bibfnamefont {M.}~\bibnamefont {Taniguchi}}, \bibinfo {author} {\bibfnamefont {H.}~\bibnamefont {Negishi}}, \bibinfo {author} {\bibfnamefont {T.}~\bibnamefont {Oguchi}}, \bibinfo {author} {\bibfnamefont {T.}~\bibnamefont {Saitoh}}, \bibinfo {author} {\bibfnamefont {A.}~\bibnamefont {Fujimori}},\ and\ \bibinfo {author} {\bibfnamefont {T.}~\bibnamefont {Kanomata}},\ }\bibfield  {title} {\bibinfo {title} {{Angle-resolved photoemission study of the strongly correlated semiconductor FeSi}},\ }\href {https://doi.org/10.1103/PhysRevB.77.205117} {\bibfield  {journal} {\bibinfo  {journal} {Phys. Rev. B}\ }\textbf {\bibinfo {volume} {77}},\ \bibinfo {pages} {205117}
  (\bibinfo {year} {2008})}\BibitemShut {NoStop}%
\bibitem [{\citenamefont {Khan}\ \emph {et~al.}(2022)\citenamefont {Khan}, \citenamefont {Krannich}, \citenamefont {Boll}, \citenamefont {Heid}, \citenamefont {Lamago}, \citenamefont {Ivanov}, \citenamefont {Voneshen},\ and\ \citenamefont {Weber}}]{Kahn2022PRBFeSiINSPhononSoftening}%
  \BibitemOpen
  \bibfield  {author} {\bibinfo {author} {\bibfnamefont {N.}~\bibnamefont {Khan}}, \bibinfo {author} {\bibfnamefont {S.}~\bibnamefont {Krannich}}, \bibinfo {author} {\bibfnamefont {D.}~\bibnamefont {Boll}}, \bibinfo {author} {\bibfnamefont {R.}~\bibnamefont {Heid}}, \bibinfo {author} {\bibfnamefont {D.}~\bibnamefont {Lamago}}, \bibinfo {author} {\bibfnamefont {A.}~\bibnamefont {Ivanov}}, \bibinfo {author} {\bibfnamefont {D.}~\bibnamefont {Voneshen}},\ and\ \bibinfo {author} {\bibfnamefont {F.}~\bibnamefont {Weber}},\ }\bibfield  {title} {\bibinfo {title} {Combined inelastic neutron scattering and ab initio lattice dynamics study of fesi},\ }\href {https://doi.org/10.1103/PhysRevB.105.134304} {\bibfield  {journal} {\bibinfo  {journal} {Phys. Rev. B}\ }\textbf {\bibinfo {volume} {105}},\ \bibinfo {pages} {134304} (\bibinfo {year} {2022})}\BibitemShut {NoStop}%
\bibitem [{\citenamefont {Jaccarino}\ \emph {et~al.}(1967)\citenamefont {Jaccarino}, \citenamefont {Wertheim}, \citenamefont {Wernick}, \citenamefont {Walker},\ and\ \citenamefont {Arajs}}]{Jaccarino1967PhysRev.160.476EarlyFeSiMagnetism}%
  \BibitemOpen
  \bibfield  {author} {\bibinfo {author} {\bibfnamefont {V.}~\bibnamefont {Jaccarino}}, \bibinfo {author} {\bibfnamefont {G.~K.}\ \bibnamefont {Wertheim}}, \bibinfo {author} {\bibfnamefont {J.~H.}\ \bibnamefont {Wernick}}, \bibinfo {author} {\bibfnamefont {L.~R.}\ \bibnamefont {Walker}},\ and\ \bibinfo {author} {\bibfnamefont {S.}~\bibnamefont {Arajs}},\ }\bibfield  {title} {\bibinfo {title} {{Paramagnetic Excited State of FeSi}},\ }\href {https://doi.org/10.1103/PhysRev.160.476} {\bibfield  {journal} {\bibinfo  {journal} {Phys. Rev.}\ }\textbf {\bibinfo {volume} {160}},\ \bibinfo {pages} {476} (\bibinfo {year} {1967})}\BibitemShut {NoStop}%
\bibitem [{\citenamefont {Hunt}\ \emph {et~al.}(1994)\citenamefont {Hunt}, \citenamefont {Chernikov}, \citenamefont {Felder}, \citenamefont {Ott}, \citenamefont {Fisk},\ and\ \citenamefont {Canfield}}]{Hunt1994PRBFeSiMagThermTransSbFlux}%
  \BibitemOpen
  \bibfield  {author} {\bibinfo {author} {\bibfnamefont {M.~B.}\ \bibnamefont {Hunt}}, \bibinfo {author} {\bibfnamefont {M.~A.}\ \bibnamefont {Chernikov}}, \bibinfo {author} {\bibfnamefont {E.}~\bibnamefont {Felder}}, \bibinfo {author} {\bibfnamefont {H.~R.}\ \bibnamefont {Ott}}, \bibinfo {author} {\bibfnamefont {Z.}~\bibnamefont {Fisk}},\ and\ \bibinfo {author} {\bibfnamefont {P.}~\bibnamefont {Canfield}},\ }\bibfield  {title} {\bibinfo {title} {{Low-temperature magnetic, thermal, and transport properties of FeSi}},\ }\href {https://doi.org/10.1103/PhysRevB.50.14933} {\bibfield  {journal} {\bibinfo  {journal} {Phys. Rev. B}\ }\textbf {\bibinfo {volume} {50}},\ \bibinfo {pages} {14933} (\bibinfo {year} {1994})}\BibitemShut {NoStop}%
\bibitem [{\citenamefont {Fang}\ \emph {et~al.}(2018)\citenamefont {Fang}, \citenamefont {Ran}, \citenamefont {Xie}, \citenamefont {Wang}, \citenamefont {Meng},\ and\ \citenamefont {Maple}}]{fang2018evidence}%
  \BibitemOpen
  \bibfield  {author} {\bibinfo {author} {\bibfnamefont {Y.}~\bibnamefont {Fang}}, \bibinfo {author} {\bibfnamefont {S.}~\bibnamefont {Ran}}, \bibinfo {author} {\bibfnamefont {W.}~\bibnamefont {Xie}}, \bibinfo {author} {\bibfnamefont {S.}~\bibnamefont {Wang}}, \bibinfo {author} {\bibfnamefont {Y.~S.}\ \bibnamefont {Meng}},\ and\ \bibinfo {author} {\bibfnamefont {M.~B.}\ \bibnamefont {Maple}},\ }\bibfield  {title} {\bibinfo {title} {{Evidence for a conducting surface ground state in high-quality single crystalline FeSi}},\ }\href@noop {} {\bibfield  {journal} {\bibinfo  {journal} {Proceedings of the National Academy of Sciences}\ }\textbf {\bibinfo {volume} {115}},\ \bibinfo {pages} {8558} (\bibinfo {year} {2018})}\BibitemShut {NoStop}%
\bibitem [{\citenamefont {Yang}\ \emph {et~al.}(2021)\citenamefont {Yang}, \citenamefont {Uphoff}, \citenamefont {Zhang}, \citenamefont {Reichert}, \citenamefont {Seitsonen}, \citenamefont {Bauer}, \citenamefont {Pfleiderer},\ and\ \citenamefont {Barth}}]{Yang_PNAS_2021STS}%
  \BibitemOpen
  \bibfield  {author} {\bibinfo {author} {\bibfnamefont {B.}~\bibnamefont {Yang}}, \bibinfo {author} {\bibfnamefont {M.}~\bibnamefont {Uphoff}}, \bibinfo {author} {\bibfnamefont {Y.-Q.}\ \bibnamefont {Zhang}}, \bibinfo {author} {\bibfnamefont {J.}~\bibnamefont {Reichert}}, \bibinfo {author} {\bibfnamefont {A.~P.}\ \bibnamefont {Seitsonen}}, \bibinfo {author} {\bibfnamefont {A.}~\bibnamefont {Bauer}}, \bibinfo {author} {\bibfnamefont {C.}~\bibnamefont {Pfleiderer}},\ and\ \bibinfo {author} {\bibfnamefont {J.~V.}\ \bibnamefont {Barth}},\ }\bibfield  {title} {\bibinfo {title} {{Atomistic investigation of surface characteristics and electronic features at high-purity FeSi(110) presenting interfacial metallicity}},\ }\href {https://doi.org/10.1073/pnas.2021203118} {\bibfield  {journal} {\bibinfo  {journal} {Proceedings of the National Academy of Sciences}\ }\textbf {\bibinfo {volume} {118}},\ \bibinfo {pages} {e2021203118} (\bibinfo {year} {2021})},\ \Eprint
  {https://arxiv.org/abs/https://www.pnas.org/doi/pdf/10.1073/pnas.2021203118} {https://www.pnas.org/doi/pdf/10.1073/pnas.2021203118} \BibitemShut {NoStop}%
\bibitem [{\citenamefont {Zhang}\ \emph {et~al.}(2013)\citenamefont {Zhang}, \citenamefont {Butch}, \citenamefont {Syers}, \citenamefont {Ziemak}, \citenamefont {Greene},\ and\ \citenamefont {Paglione}}]{Zhang2012PhysRevX.3.011011SmB6}%
  \BibitemOpen
  \bibfield  {author} {\bibinfo {author} {\bibfnamefont {X.}~\bibnamefont {Zhang}}, \bibinfo {author} {\bibfnamefont {N.~P.}\ \bibnamefont {Butch}}, \bibinfo {author} {\bibfnamefont {P.}~\bibnamefont {Syers}}, \bibinfo {author} {\bibfnamefont {S.}~\bibnamefont {Ziemak}}, \bibinfo {author} {\bibfnamefont {R.~L.}\ \bibnamefont {Greene}},\ and\ \bibinfo {author} {\bibfnamefont {J.}~\bibnamefont {Paglione}},\ }\bibfield  {title} {\bibinfo {title} {{Hybridization, Inter-Ion Correlation, and Surface States in the Kondo Insulator ${\mathrm{SmB}}_{6}$}},\ }\href {https://doi.org/10.1103/PhysRevX.3.011011} {\bibfield  {journal} {\bibinfo  {journal} {Phys. Rev. X}\ }\textbf {\bibinfo {volume} {3}},\ \bibinfo {pages} {011011} (\bibinfo {year} {2013})}\BibitemShut {NoStop}%
\bibitem [{\citenamefont {Zhu}\ \emph {et~al.}(2013)\citenamefont {Zhu}, \citenamefont {Nicolaou}, \citenamefont {Levy}, \citenamefont {Butch}, \citenamefont {Syers}, \citenamefont {Wang}, \citenamefont {Paglione}, \citenamefont {Sawatzky}, \citenamefont {Elfimov},\ and\ \citenamefont {Damascelli}}]{Zhu2013PhysRevLett.111.216402SmB6danglingBonds}%
  \BibitemOpen
  \bibfield  {author} {\bibinfo {author} {\bibfnamefont {Z.-H.}\ \bibnamefont {Zhu}}, \bibinfo {author} {\bibfnamefont {A.}~\bibnamefont {Nicolaou}}, \bibinfo {author} {\bibfnamefont {G.}~\bibnamefont {Levy}}, \bibinfo {author} {\bibfnamefont {N.~P.}\ \bibnamefont {Butch}}, \bibinfo {author} {\bibfnamefont {P.}~\bibnamefont {Syers}}, \bibinfo {author} {\bibfnamefont {X.~F.}\ \bibnamefont {Wang}}, \bibinfo {author} {\bibfnamefont {J.}~\bibnamefont {Paglione}}, \bibinfo {author} {\bibfnamefont {G.~A.}\ \bibnamefont {Sawatzky}}, \bibinfo {author} {\bibfnamefont {I.~S.}\ \bibnamefont {Elfimov}},\ and\ \bibinfo {author} {\bibfnamefont {A.}~\bibnamefont {Damascelli}},\ }\bibfield  {title} {\bibinfo {title} {{Polarity-Driven Surface Metallicity in ${\mathrm{SmB}}_{6}$}},\ }\href {https://doi.org/10.1103/PhysRevLett.111.216402} {\bibfield  {journal} {\bibinfo  {journal} {Phys. Rev. Lett.}\ }\textbf {\bibinfo {volume} {111}},\ \bibinfo {pages} {216402} (\bibinfo {year} {2013})}\BibitemShut {NoStop}%
\bibitem [{\citenamefont {Eo}\ \emph {et~al.}(2023)\citenamefont {Eo}, \citenamefont {Avers}, \citenamefont {Horn}, \citenamefont {Yoon}, \citenamefont {Saha}, \citenamefont {Suarez}, \citenamefont {Fuhrer},\ and\ \citenamefont {Paglione}}]{Eo2023APL_FeSiFeSb2}%
  \BibitemOpen
  \bibfield  {author} {\bibinfo {author} {\bibfnamefont {Y.~S.}\ \bibnamefont {Eo}}, \bibinfo {author} {\bibfnamefont {K.}~\bibnamefont {Avers}}, \bibinfo {author} {\bibfnamefont {J.~A.}\ \bibnamefont {Horn}}, \bibinfo {author} {\bibfnamefont {H.}~\bibnamefont {Yoon}}, \bibinfo {author} {\bibfnamefont {S.~R.}\ \bibnamefont {Saha}}, \bibinfo {author} {\bibfnamefont {A.}~\bibnamefont {Suarez}}, \bibinfo {author} {\bibfnamefont {M.~S.}\ \bibnamefont {Fuhrer}},\ and\ \bibinfo {author} {\bibfnamefont {J.}~\bibnamefont {Paglione}},\ }\bibfield  {title} {\bibinfo {title} {{Extraordinary bulk-insulating behavior in the strongly correlated materials FeSi and FeSb$_2$}},\ }\href {https://doi.org/10.1063/5.0148249} {\bibfield  {journal} {\bibinfo  {journal} {Applied Physics Letters}\ }\textbf {\bibinfo {volume} {122}},\ \bibinfo {pages} {233102} (\bibinfo {year} {2023})}\BibitemShut {NoStop}%
\bibitem [{\citenamefont {Sarma}\ and\ \citenamefont {Hwang}(2014)}]{MIT_DasSarma}%
  \BibitemOpen
  \bibfield  {author} {\bibinfo {author} {\bibfnamefont {S.~D.}\ \bibnamefont {Sarma}}\ and\ \bibinfo {author} {\bibfnamefont {E.~H.}\ \bibnamefont {Hwang}},\ }\bibfield  {title} {\bibinfo {title} {Two-dimensional metal-insulator transition as a strong localization induced crossover phenomenon},\ }\href {https://doi.org/10.1103/PhysRevB.89.235423} {\bibfield  {journal} {\bibinfo  {journal} {Phys. Rev. B}\ }\textbf {\bibinfo {volume} {89}},\ \bibinfo {pages} {235423} (\bibinfo {year} {2014})}\BibitemShut {NoStop}%
\bibitem [{\citenamefont {Zabrodskii}(2001)}]{Zabrodskii2001PhilMagBColumbgap}%
  \BibitemOpen
  \bibfield  {author} {\bibinfo {author} {\bibfnamefont {A.~G.}\ \bibnamefont {Zabrodskii}},\ }\bibfield  {title} {\bibinfo {title} {The coulomb gap: The view of an experimenter},\ }\href {https://doi.org/10.1080/13642810108205796} {\bibfield  {journal} {\bibinfo  {journal} {Philosophical Magazine B}\ }\textbf {\bibinfo {volume} {81}},\ \bibinfo {pages} {1131} (\bibinfo {year} {2001})},\ \Eprint {https://arxiv.org/abs/https://doi.org/10.1080/13642810108205796} {https://doi.org/10.1080/13642810108205796} \BibitemShut {NoStop}%
\bibitem [{\citenamefont {Sarma}\ and\ \citenamefont {Hwang}(2005)}]{SARMA2005SSComm2DMITransition}%
  \BibitemOpen
  \bibfield  {author} {\bibinfo {author} {\bibfnamefont {S.~D.}\ \bibnamefont {Sarma}}\ and\ \bibinfo {author} {\bibfnamefont {E.}~\bibnamefont {Hwang}},\ }\bibfield  {title} {\bibinfo {title} {{The so-called two dimensional metal–insulator transition}},\ }\href {https://doi.org/https://doi.org/10.1016/j.ssc.2005.04.035} {\bibfield  {journal} {\bibinfo  {journal} {Solid State Communications}\ }\textbf {\bibinfo {volume} {135}},\ \bibinfo {pages} {579} (\bibinfo {year} {2005})},\ \bibinfo {note} {fundamental Optical and Quantum Effects in Condensed Matter}\BibitemShut {NoStop}%
\bibitem [{\citenamefont {Huang}\ \emph {et~al.}(2022{\natexlab{a}})\citenamefont {Huang}, \citenamefont {Skinner},\ and\ \citenamefont {Shklovskii}}]{Huang_2022_2DTopInsMetalTrans}%
  \BibitemOpen
  \bibfield  {author} {\bibinfo {author} {\bibfnamefont {Y.}~\bibnamefont {Huang}}, \bibinfo {author} {\bibfnamefont {B.}~\bibnamefont {Skinner}},\ and\ \bibinfo {author} {\bibfnamefont {B.~I.}\ \bibnamefont {Shklovskii}},\ }\bibfield  {title} {\bibinfo {title} {{Conductivity of Two-Dimensional Small Gap Semiconductors and Topological Insulators in Strong Coulomb Disorder}},\ }\href {https://doi.org/https://doi.org/10.1134/S1063776122100065} {\bibfield  {journal} {\bibinfo  {journal} {Journal of Experimental and Theoretical Physics}\ }\textbf {\bibinfo {volume} {135}},\ \bibinfo {pages} {409} (\bibinfo {year} {2022}{\natexlab{a}})}\BibitemShut {NoStop}%
\bibitem [{\citenamefont {Huang}\ \emph {et~al.}(2022{\natexlab{b}})\citenamefont {Huang}, \citenamefont {He}, \citenamefont {Skinner},\ and\ \citenamefont {Shklovskii}}]{Huang2022PhysRevB.105.054206CoulombDisorder2Dsemiconductors}%
  \BibitemOpen
  \bibfield  {author} {\bibinfo {author} {\bibfnamefont {Y.}~\bibnamefont {Huang}}, \bibinfo {author} {\bibfnamefont {Y.}~\bibnamefont {He}}, \bibinfo {author} {\bibfnamefont {B.}~\bibnamefont {Skinner}},\ and\ \bibinfo {author} {\bibfnamefont {B.~I.}\ \bibnamefont {Shklovskii}},\ }\bibfield  {title} {\bibinfo {title} {{Conductivity of two-dimensional narrow gap semiconductors subjected to strong Coulomb disorder}},\ }\href {https://doi.org/10.1103/PhysRevB.105.054206} {\bibfield  {journal} {\bibinfo  {journal} {Phys. Rev. B}\ }\textbf {\bibinfo {volume} {105}},\ \bibinfo {pages} {054206} (\bibinfo {year} {2022}{\natexlab{b}})}\BibitemShut {NoStop}%
\bibitem [{\citenamefont {Schlesinger}\ \emph {et~al.}(1993)\citenamefont {Schlesinger}, \citenamefont {Fisk}, \citenamefont {Zhang}, \citenamefont {Maple}, \citenamefont {DiTusa},\ and\ \citenamefont {Aeppli}}]{FeSi_Schlesinger}%
  \BibitemOpen
  \bibfield  {author} {\bibinfo {author} {\bibfnamefont {Z.}~\bibnamefont {Schlesinger}}, \bibinfo {author} {\bibfnamefont {Z.}~\bibnamefont {Fisk}}, \bibinfo {author} {\bibfnamefont {H.-T.}\ \bibnamefont {Zhang}}, \bibinfo {author} {\bibfnamefont {M.~B.}\ \bibnamefont {Maple}}, \bibinfo {author} {\bibfnamefont {J.}~\bibnamefont {DiTusa}},\ and\ \bibinfo {author} {\bibfnamefont {G.}~\bibnamefont {Aeppli}},\ }\bibfield  {title} {\bibinfo {title} {{Unconventional charge gap formation in FeSi}},\ }\href {https://doi.org/10.1103/PhysRevLett.71.1748} {\bibfield  {journal} {\bibinfo  {journal} {Phys. Rev. Lett.}\ }\textbf {\bibinfo {volume} {71}},\ \bibinfo {pages} {1748} (\bibinfo {year} {1993})}\BibitemShut {NoStop}%
\bibitem [{\citenamefont {Buschinger}\ \emph {et~al.}(1997)\citenamefont {Buschinger}, \citenamefont {Geibel}, \citenamefont {Steglich}, \citenamefont {Mandrus}, \citenamefont {Young}, \citenamefont {Sarrao},\ and\ \citenamefont {Fisk}}]{FeSiOldElecAndThermCond_BUSCHINGER}%
  \BibitemOpen
  \bibfield  {author} {\bibinfo {author} {\bibfnamefont {B.}~\bibnamefont {Buschinger}}, \bibinfo {author} {\bibfnamefont {C.}~\bibnamefont {Geibel}}, \bibinfo {author} {\bibfnamefont {F.}~\bibnamefont {Steglich}}, \bibinfo {author} {\bibfnamefont {D.}~\bibnamefont {Mandrus}}, \bibinfo {author} {\bibfnamefont {D.}~\bibnamefont {Young}}, \bibinfo {author} {\bibfnamefont {J.}~\bibnamefont {Sarrao}},\ and\ \bibinfo {author} {\bibfnamefont {Z.}~\bibnamefont {Fisk}},\ }\bibfield  {title} {\bibinfo {title} {Transport properties of fesi},\ }\href {https://doi.org/https://doi.org/10.1016/S0921-4526(96)00839-3} {\bibfield  {journal} {\bibinfo  {journal} {Physica B: Condensed Matter}\ }\textbf {\bibinfo {volume} {230-232}},\ \bibinfo {pages} {784} (\bibinfo {year} {1997})},\ \bibinfo {note} {proceedings of the International Conference on Strongly Correlated Electron Systems}\BibitemShut {NoStop}%
\bibitem [{\citenamefont {Degiorgi}\ \emph {et~al.}(1995{\natexlab{a}})\citenamefont {Degiorgi}, \citenamefont {Hunt}, \citenamefont {Ott},\ and\ \citenamefont {Fisk}}]{SbFluxFeSi_Degiorgi}%
  \BibitemOpen
  \bibfield  {author} {\bibinfo {author} {\bibfnamefont {L.}~\bibnamefont {Degiorgi}}, \bibinfo {author} {\bibfnamefont {M.}~\bibnamefont {Hunt}}, \bibinfo {author} {\bibfnamefont {H.~R.}\ \bibnamefont {Ott}},\ and\ \bibinfo {author} {\bibfnamefont {Z.}~\bibnamefont {Fisk}},\ }\bibfield  {title} {\bibinfo {title} {{Transport and optical properties of FeSi}},\ }\href {https://doi.org/10.1016/0921-4526(94)00592-J} {\bibfield  {journal} {\bibinfo  {journal} {PHYSICA B-CONDENSED MATTER}\ }\textbf {\bibinfo {volume} {206}},\ \bibinfo {pages} {810} (\bibinfo {year} {1995}{\natexlab{a}})}\BibitemShut {NoStop}%
\bibitem [{\citenamefont {Glushkov}\ \emph {et~al.}(2000)\citenamefont {Glushkov}, \citenamefont {Sluchanko}, \citenamefont {Demishev}, \citenamefont {Kondrin}, \citenamefont {Pronin}, \citenamefont {Petukhov}, \citenamefont {Bruynseraede}, \citenamefont {Moshchalkov},\ and\ \citenamefont {Menovsky}}]{FeSiTransAnom_Glushkov}%
  \BibitemOpen
  \bibfield  {author} {\bibinfo {author} {\bibfnamefont {V.}~\bibnamefont {Glushkov}}, \bibinfo {author} {\bibfnamefont {N.}~\bibnamefont {Sluchanko}}, \bibinfo {author} {\bibfnamefont {S.}~\bibnamefont {Demishev}}, \bibinfo {author} {\bibfnamefont {M.}~\bibnamefont {Kondrin}}, \bibinfo {author} {\bibfnamefont {A.}~\bibnamefont {Pronin}}, \bibinfo {author} {\bibfnamefont {K.}~\bibnamefont {Petukhov}}, \bibinfo {author} {\bibfnamefont {Y.}~\bibnamefont {Bruynseraede}}, \bibinfo {author} {\bibfnamefont {V.}~\bibnamefont {Moshchalkov}},\ and\ \bibinfo {author} {\bibfnamefont {A.}~\bibnamefont {Menovsky}},\ }\bibfield  {title} {\bibinfo {title} {{Low-temperature transport anomalies in FeSi}},\ }\href {https://doi.org/10.1016/S0921-4526(99)02605-8} {\bibfield  {journal} {\bibinfo  {journal} {PHYSICA B}\ }\textbf {\bibinfo {volume} {284}},\ \bibinfo {pages} {1179} (\bibinfo {year} {2000})},\ \bibinfo {note} {22nd International Conference on Low Temperature Physics, HELSINKI UNIV TECHNOL, HELSINKI, FINLAND, AUG
  04-11, 1999}\BibitemShut {NoStop}%
\bibitem [{\citenamefont {Lisunov}\ \emph {et~al.}(1996)\citenamefont {Lisunov}, \citenamefont {Arushanov}, \citenamefont {Kloc}, \citenamefont {Broto}, \citenamefont {Leotin}, \citenamefont {Rokoto}, \citenamefont {Respaud},\ and\ \citenamefont {Bucher}}]{FeSiAndersonLocalized_Lisunov}%
  \BibitemOpen
  \bibfield  {author} {\bibinfo {author} {\bibfnamefont {K.}~\bibnamefont {Lisunov}}, \bibinfo {author} {\bibfnamefont {E.}~\bibnamefont {Arushanov}}, \bibinfo {author} {\bibfnamefont {C.}~\bibnamefont {Kloc}}, \bibinfo {author} {\bibfnamefont {J.}~\bibnamefont {Broto}}, \bibinfo {author} {\bibfnamefont {J.}~\bibnamefont {Leotin}}, \bibinfo {author} {\bibfnamefont {H.}~\bibnamefont {Rokoto}}, \bibinfo {author} {\bibfnamefont {M.}~\bibnamefont {Respaud}},\ and\ \bibinfo {author} {\bibfnamefont {E.}~\bibnamefont {Bucher}},\ }\bibfield  {title} {\bibinfo {title} {{Conductivity and magnetoresistance of FeSi in the Anderson-localized regime}},\ }\href {https://doi.org/10.1016/S0921-4526(96)00508-X} {\bibfield  {journal} {\bibinfo  {journal} {PHYSICA B}\ }\textbf {\bibinfo {volume} {229}},\ \bibinfo {pages} {37} (\bibinfo {year} {1996})}\BibitemShut {NoStop}%
\bibitem [{\citenamefont {Mihalik}\ \emph {et~al.}(1996)\citenamefont {Mihalik}, \citenamefont {Timko}, \citenamefont {Samuely}, \citenamefont {Tomašovičova-Hudákova}, \citenamefont {Szabó},\ and\ \citenamefont {Menovsky}}]{MIHALIK_1996_PointContact}%
  \BibitemOpen
  \bibfield  {author} {\bibinfo {author} {\bibfnamefont {M.}~\bibnamefont {Mihalik}}, \bibinfo {author} {\bibfnamefont {M.}~\bibnamefont {Timko}}, \bibinfo {author} {\bibfnamefont {P.}~\bibnamefont {Samuely}}, \bibinfo {author} {\bibfnamefont {N.}~\bibnamefont {Tomašovičova-Hudákova}}, \bibinfo {author} {\bibfnamefont {P.}~\bibnamefont {Szabó}},\ and\ \bibinfo {author} {\bibfnamefont {A.}~\bibnamefont {Menovsky}},\ }\bibfield  {title} {\bibinfo {title} {{Magnetic properties and gap formation in FeSi}},\ }\href {https://doi.org/https://doi.org/10.1016/0304-8853(95)01235-4} {\bibfield  {journal} {\bibinfo  {journal} {Journal of Magnetism and Magnetic Materials}\ }\textbf {\bibinfo {volume} {157-158}},\ \bibinfo {pages} {637} (\bibinfo {year} {1996})},\ \bibinfo {note} {european Magnetic Materials and Applications Conference}\BibitemShut {NoStop}%
\bibitem [{\citenamefont {Paschen}\ \emph {et~al.}(1997)\citenamefont {Paschen}, \citenamefont {Felder}, \citenamefont {Chernikov}, \citenamefont {Degiorgi}, \citenamefont {Schwer}, \citenamefont {Ott}, \citenamefont {Young}, \citenamefont {Sarrao},\ and\ \citenamefont {Fisk}}]{Paschen_1997_DilFridge}%
  \BibitemOpen
  \bibfield  {author} {\bibinfo {author} {\bibfnamefont {S.}~\bibnamefont {Paschen}}, \bibinfo {author} {\bibfnamefont {E.}~\bibnamefont {Felder}}, \bibinfo {author} {\bibfnamefont {M.~A.}\ \bibnamefont {Chernikov}}, \bibinfo {author} {\bibfnamefont {L.}~\bibnamefont {Degiorgi}}, \bibinfo {author} {\bibfnamefont {H.}~\bibnamefont {Schwer}}, \bibinfo {author} {\bibfnamefont {H.~R.}\ \bibnamefont {Ott}}, \bibinfo {author} {\bibfnamefont {D.~P.}\ \bibnamefont {Young}}, \bibinfo {author} {\bibfnamefont {J.~L.}\ \bibnamefont {Sarrao}},\ and\ \bibinfo {author} {\bibfnamefont {Z.}~\bibnamefont {Fisk}},\ }\bibfield  {title} {\bibinfo {title} {{Low-temperature transport, thermodynamic, and optical properties of FeSi}},\ }\href {https://doi.org/10.1103/PhysRevB.56.12916} {\bibfield  {journal} {\bibinfo  {journal} {Phys. Rev. B}\ }\textbf {\bibinfo {volume} {56}},\ \bibinfo {pages} {12916} (\bibinfo {year} {1997})}\BibitemShut {NoStop}%
\bibitem [{\citenamefont {Changdar}\ \emph {et~al.}(2020)\citenamefont {Changdar}, \citenamefont {Aswartham}, \citenamefont {Bose}, \citenamefont {Kushnirenko}, \citenamefont {Shipunov}, \citenamefont {Plumb}, \citenamefont {Shi}, \citenamefont {Narayan}, \citenamefont {B\"uchner},\ and\ \citenamefont {Thirupathaiah}}]{Changdar_2020_PRBARPESchiral}%
  \BibitemOpen
  \bibfield  {author} {\bibinfo {author} {\bibfnamefont {S.}~\bibnamefont {Changdar}}, \bibinfo {author} {\bibfnamefont {S.}~\bibnamefont {Aswartham}}, \bibinfo {author} {\bibfnamefont {A.}~\bibnamefont {Bose}}, \bibinfo {author} {\bibfnamefont {Y.}~\bibnamefont {Kushnirenko}}, \bibinfo {author} {\bibfnamefont {G.}~\bibnamefont {Shipunov}}, \bibinfo {author} {\bibfnamefont {N.~C.}\ \bibnamefont {Plumb}}, \bibinfo {author} {\bibfnamefont {M.}~\bibnamefont {Shi}}, \bibinfo {author} {\bibfnamefont {A.}~\bibnamefont {Narayan}}, \bibinfo {author} {\bibfnamefont {B.}~\bibnamefont {B\"uchner}},\ and\ \bibinfo {author} {\bibfnamefont {S.}~\bibnamefont {Thirupathaiah}},\ }\bibfield  {title} {\bibinfo {title} {{Electronic structure studies of FeSi: A chiral topological system}},\ }\href {https://doi.org/10.1103/PhysRevB.101.235105} {\bibfield  {journal} {\bibinfo  {journal} {Phys. Rev. B}\ }\textbf {\bibinfo {volume} {101}},\ \bibinfo {pages} {235105} (\bibinfo {year} {2020})}\BibitemShut {NoStop}%
\bibitem [{\citenamefont {Lunkenheimer}\ \emph {et~al.}(1995)\citenamefont {Lunkenheimer}, \citenamefont {Knebel}, \citenamefont {Viana},\ and\ \citenamefont {Loidl}}]{LUNKENHEIMER_1995_Hopping}%
  \BibitemOpen
  \bibfield  {author} {\bibinfo {author} {\bibfnamefont {P.}~\bibnamefont {Lunkenheimer}}, \bibinfo {author} {\bibfnamefont {G.}~\bibnamefont {Knebel}}, \bibinfo {author} {\bibfnamefont {R.}~\bibnamefont {Viana}},\ and\ \bibinfo {author} {\bibfnamefont {A.}~\bibnamefont {Loidl}},\ }\bibfield  {title} {\bibinfo {title} {{Hopping conductivity in FeSi}},\ }\href {https://doi.org/https://doi.org/10.1016/0038-1098(94)00895-7} {\bibfield  {journal} {\bibinfo  {journal} {Solid State Communications}\ }\textbf {\bibinfo {volume} {93}},\ \bibinfo {pages} {891} (\bibinfo {year} {1995})}\BibitemShut {NoStop}%
\bibitem [{\citenamefont {Degiorgi}\ \emph {et~al.}(1995{\natexlab{b}})\citenamefont {Degiorgi}, \citenamefont {Hunt}, \citenamefont {Ott},\ and\ \citenamefont {Fisk}}]{DEGIORGI_1995_electrodynamic}%
  \BibitemOpen
  \bibfield  {author} {\bibinfo {author} {\bibfnamefont {L.}~\bibnamefont {Degiorgi}}, \bibinfo {author} {\bibfnamefont {M.}~\bibnamefont {Hunt}}, \bibinfo {author} {\bibfnamefont {H.}~\bibnamefont {Ott}},\ and\ \bibinfo {author} {\bibfnamefont {Z.}~\bibnamefont {Fisk}},\ }\bibfield  {title} {\bibinfo {title} {{Transport and optical properties of FeSi}},\ }\href {https://doi.org/https://doi.org/10.1016/0921-4526(94)00592-J} {\bibfield  {journal} {\bibinfo  {journal} {Physica B: Condensed Matter}\ }\textbf {\bibinfo {volume} {206-207}},\ \bibinfo {pages} {810} (\bibinfo {year} {1995}{\natexlab{b}})},\ \bibinfo {note} {proceedings of the International Conference on Strongly Correlated Electron Systems}\BibitemShut {NoStop}%
\bibitem [{\citenamefont {Ou-Yang}\ \emph {et~al.}(2017)\citenamefont {Ou-Yang}, \citenamefont {Zhuang}, \citenamefont {Ramachandran}, \citenamefont {Chen}, \citenamefont {Shu}, \citenamefont {Hu}, \citenamefont {Chou},\ and\ \citenamefont {Kuo}}]{OUYANG_2017_CoSub}%
  \BibitemOpen
  \bibfield  {author} {\bibinfo {author} {\bibfnamefont {T.}~\bibnamefont {Ou-Yang}}, \bibinfo {author} {\bibfnamefont {Y.}~\bibnamefont {Zhuang}}, \bibinfo {author} {\bibfnamefont {B.}~\bibnamefont {Ramachandran}}, \bibinfo {author} {\bibfnamefont {W.}~\bibnamefont {Chen}}, \bibinfo {author} {\bibfnamefont {G.}~\bibnamefont {Shu}}, \bibinfo {author} {\bibfnamefont {C.}~\bibnamefont {Hu}}, \bibinfo {author} {\bibfnamefont {F.}~\bibnamefont {Chou}},\ and\ \bibinfo {author} {\bibfnamefont {Y.}~\bibnamefont {Kuo}},\ }\bibfield  {title} {\bibinfo {title} {{Effect of Co substitution on thermoelectric properties of FeSi}},\ }\href {https://doi.org/https://doi.org/10.1016/j.jallcom.2017.01.217} {\bibfield  {journal} {\bibinfo  {journal} {Journal of Alloys and Compounds}\ }\textbf {\bibinfo {volume} {702}},\ \bibinfo {pages} {92} (\bibinfo {year} {2017})}\BibitemShut {NoStop}%
\bibitem [{\citenamefont {Delaire}\ \emph {et~al.}(2015)\citenamefont {Delaire}, \citenamefont {Al-Qasir}, \citenamefont {May}, \citenamefont {Li}, \citenamefont {Sales}, \citenamefont {Niedziela}, \citenamefont {Ma}, \citenamefont {Matsuda}, \citenamefont {Abernathy},\ and\ \citenamefont {Berlijn}}]{Delaire2015PhysRevB.91.094307IrOsDopping}%
  \BibitemOpen
  \bibfield  {author} {\bibinfo {author} {\bibfnamefont {O.}~\bibnamefont {Delaire}}, \bibinfo {author} {\bibfnamefont {I.~I.}\ \bibnamefont {Al-Qasir}}, \bibinfo {author} {\bibfnamefont {A.~F.}\ \bibnamefont {May}}, \bibinfo {author} {\bibfnamefont {C.~W.}\ \bibnamefont {Li}}, \bibinfo {author} {\bibfnamefont {B.~C.}\ \bibnamefont {Sales}}, \bibinfo {author} {\bibfnamefont {J.~L.}\ \bibnamefont {Niedziela}}, \bibinfo {author} {\bibfnamefont {J.}~\bibnamefont {Ma}}, \bibinfo {author} {\bibfnamefont {M.}~\bibnamefont {Matsuda}}, \bibinfo {author} {\bibfnamefont {D.~L.}\ \bibnamefont {Abernathy}},\ and\ \bibinfo {author} {\bibfnamefont {T.}~\bibnamefont {Berlijn}},\ }\bibfield  {title} {\bibinfo {title} {{Heavy-impurity resonance, hybridization, and phonon spectral functions in ${\text{Fe}}_{1\ensuremath{-}x}{M}_{x}\text{Si} (M=\mathrm{Ir}, \mathrm{Os})$}},\ }\href {https://doi.org/10.1103/PhysRevB.91.094307} {\bibfield  {journal} {\bibinfo  {journal} {Phys. Rev. B}\ }\textbf {\bibinfo {volume} {91}},\ \bibinfo
  {pages} {094307} (\bibinfo {year} {2015})}\BibitemShut {NoStop}%
\bibitem [{\citenamefont {Ángel R~Ruiz}\ \emph {et~al.}(2018)\citenamefont {Ángel R~Ruiz}, \citenamefont {Hernández-Pérez}, \citenamefont {Fonseca}, \citenamefont {José-Yacamán}, \citenamefont {Ortega},\ and\ \citenamefont {Ponce}}]{Ruiz_2019NanotechFeSiNanowires}%
  \BibitemOpen
  \bibfield  {author} {\bibinfo {author} {\bibnamefont {Ángel R~Ruiz}}, \bibinfo {author} {\bibfnamefont {J.}~\bibnamefont {Hernández-Pérez}}, \bibinfo {author} {\bibfnamefont {L.~F.}\ \bibnamefont {Fonseca}}, \bibinfo {author} {\bibfnamefont {M.}~\bibnamefont {José-Yacamán}}, \bibinfo {author} {\bibfnamefont {E.}~\bibnamefont {Ortega}},\ and\ \bibinfo {author} {\bibfnamefont {A.}~\bibnamefont {Ponce}},\ }\bibfield  {title} {\bibinfo {title} {{Single nanowire measurements of room temperature ferromagnetism in FeSi nanowires and the effects of Mn-doping}},\ }\href {https://doi.org/10.1088/1361-6528/aae5cc} {\bibfield  {journal} {\bibinfo  {journal} {Nanotechnology}\ }\textbf {\bibinfo {volume} {30}},\ \bibinfo {pages} {014001} (\bibinfo {year} {2018})}\BibitemShut {NoStop}%
\bibitem [{\citenamefont {Ohtsuka}\ \emph {et~al.}(2021)\citenamefont {Ohtsuka}, \citenamefont {Kanazawa}, \citenamefont {Hirayama}, \citenamefont {Matsui}, \citenamefont {Nomoto}, \citenamefont {Arita}, \citenamefont {Nakajima}, \citenamefont {Hanashima}, \citenamefont {Ukleev}, \citenamefont {Aoki}, \citenamefont {Mogi}, \citenamefont {Fujiwara}, \citenamefont {Tsukazaki}, \citenamefont {Ichikawa}, \citenamefont {Kawasaki},\ and\ \citenamefont {Tokura}}]{ThinFeSiZak_Ohtsuka}%
  \BibitemOpen
  \bibfield  {author} {\bibinfo {author} {\bibfnamefont {Y.}~\bibnamefont {Ohtsuka}}, \bibinfo {author} {\bibfnamefont {N.}~\bibnamefont {Kanazawa}}, \bibinfo {author} {\bibfnamefont {M.}~\bibnamefont {Hirayama}}, \bibinfo {author} {\bibfnamefont {A.}~\bibnamefont {Matsui}}, \bibinfo {author} {\bibfnamefont {T.}~\bibnamefont {Nomoto}}, \bibinfo {author} {\bibfnamefont {R.}~\bibnamefont {Arita}}, \bibinfo {author} {\bibfnamefont {T.}~\bibnamefont {Nakajima}}, \bibinfo {author} {\bibfnamefont {T.}~\bibnamefont {Hanashima}}, \bibinfo {author} {\bibfnamefont {V.}~\bibnamefont {Ukleev}}, \bibinfo {author} {\bibfnamefont {H.}~\bibnamefont {Aoki}}, \bibinfo {author} {\bibfnamefont {M.}~\bibnamefont {Mogi}}, \bibinfo {author} {\bibfnamefont {K.}~\bibnamefont {Fujiwara}}, \bibinfo {author} {\bibfnamefont {A.}~\bibnamefont {Tsukazaki}}, \bibinfo {author} {\bibfnamefont {M.}~\bibnamefont {Ichikawa}}, \bibinfo {author} {\bibfnamefont {M.}~\bibnamefont {Kawasaki}},\ and\ \bibinfo {author} {\bibfnamefont {Y.}~\bibnamefont
  {Tokura}},\ }\bibfield  {title} {\bibinfo {title} {{Emergence of spin-orbit coupled ferromagnetic surface state derived from Zak phase in a nonmagnetic insulator FeSi}},\ }\href {https://doi.org/10.1126/sciadv.abj0498} {\bibfield  {journal} {\bibinfo  {journal} {Science Advances}\ }\textbf {\bibinfo {volume} {7}},\ \bibinfo {pages} {eabj0498} (\bibinfo {year} {2021})}\BibitemShut {NoStop}%
\bibitem [{\citenamefont {Burmistrov}\ and\ \citenamefont {Repin}(2018)}]{Burmistrov2018PRB_quantCorr2DMagImp}%
  \BibitemOpen
  \bibfield  {author} {\bibinfo {author} {\bibfnamefont {I.~S.}\ \bibnamefont {Burmistrov}}\ and\ \bibinfo {author} {\bibfnamefont {E.~V.}\ \bibnamefont {Repin}},\ }\bibfield  {title} {\bibinfo {title} {{Quantum corrections to conductivity of disordered electrons due to inelastic scattering off magnetic impurities}},\ }\href {https://doi.org/10.1103/PhysRevB.98.045414} {\bibfield  {journal} {\bibinfo  {journal} {Phys. Rev. B}\ }\textbf {\bibinfo {volume} {98}},\ \bibinfo {pages} {045414} (\bibinfo {year} {2018})}\BibitemShut {NoStop}%
\bibitem [{\citenamefont {Mitra}\ \emph {et~al.}(2007)\citenamefont {Mitra}, \citenamefont {Misra}, \citenamefont {Hebard}, \citenamefont {Muttalib},\ and\ \citenamefont {W\"olfle}}]{Mitra2007PRLWeakLocalFeFilm}%
  \BibitemOpen
  \bibfield  {author} {\bibinfo {author} {\bibfnamefont {P.}~\bibnamefont {Mitra}}, \bibinfo {author} {\bibfnamefont {R.}~\bibnamefont {Misra}}, \bibinfo {author} {\bibfnamefont {A.~F.}\ \bibnamefont {Hebard}}, \bibinfo {author} {\bibfnamefont {K.~A.}\ \bibnamefont {Muttalib}},\ and\ \bibinfo {author} {\bibfnamefont {P.}~\bibnamefont {W\"olfle}},\ }\bibfield  {title} {\bibinfo {title} {{Weak-Localization Correction to the Anomalous Hall Effect in Polycrystalline Fe Films}},\ }\href {https://doi.org/10.1103/PhysRevLett.99.046804} {\bibfield  {journal} {\bibinfo  {journal} {Phys. Rev. Lett.}\ }\textbf {\bibinfo {volume} {99}},\ \bibinfo {pages} {046804} (\bibinfo {year} {2007})}\BibitemShut {NoStop}%
\bibitem [{\citenamefont {Lin}\ \emph {et~al.}(1996)\citenamefont {Lin}, \citenamefont {Novet}, \citenamefont {Johnson},\ and\ \citenamefont {Valles}}]{Lin1996PRB_MR_and_AHE_FeSiLayer}%
  \BibitemOpen
  \bibfield  {author} {\bibinfo {author} {\bibfnamefont {Y.~K.}\ \bibnamefont {Lin}}, \bibinfo {author} {\bibfnamefont {T.~R.}\ \bibnamefont {Novet}}, \bibinfo {author} {\bibfnamefont {D.~C.}\ \bibnamefont {Johnson}},\ and\ \bibinfo {author} {\bibfnamefont {J.~M.}\ \bibnamefont {Valles}},\ }\bibfield  {title} {\bibinfo {title} {{Magnetotransport studies of strongly disordered annealed amorphous Fe/Si multilayers}},\ }\href {https://doi.org/10.1103/PhysRevB.53.4796} {\bibfield  {journal} {\bibinfo  {journal} {Phys. Rev. B}\ }\textbf {\bibinfo {volume} {53}},\ \bibinfo {pages} {4796} (\bibinfo {year} {1996})}\BibitemShut {NoStop}%
\bibitem [{\citenamefont {Mermin}\ and\ \citenamefont {Wagner}(1966)}]{Mermin1966PhysRevLett.17.1133OGMWtheroem}%
  \BibitemOpen
  \bibfield  {author} {\bibinfo {author} {\bibfnamefont {N.~D.}\ \bibnamefont {Mermin}}\ and\ \bibinfo {author} {\bibfnamefont {H.}~\bibnamefont {Wagner}},\ }\bibfield  {title} {\bibinfo {title} {{Absence of Ferromagnetism or Antiferromagnetism in One- or Two-Dimensional Isotropic Heisenberg Models}},\ }\href {https://doi.org/10.1103/PhysRevLett.17.1133} {\bibfield  {journal} {\bibinfo  {journal} {Phys. Rev. Lett.}\ }\textbf {\bibinfo {volume} {17}},\ \bibinfo {pages} {1133} (\bibinfo {year} {1966})}\BibitemShut {NoStop}%
\bibitem [{\citenamefont {Gong}\ \emph {et~al.}(2017)\citenamefont {Gong}, \citenamefont {Lin}, \citenamefont {Zhenglu}, \citenamefont {Huiwen}, \citenamefont {Stern}, \citenamefont {Yang}, \citenamefont {Ting}, \citenamefont {Wei}, \citenamefont {Wang}, \citenamefont {Wang}, \citenamefont {Qui}, \citenamefont {Cava}, \citenamefont {Louie}, \citenamefont {Jing},\ and\ \citenamefont {Xiang}}]{gong2017CrGeTevDW2Dferromagnetism}%
  \BibitemOpen
  \bibfield  {author} {\bibinfo {author} {\bibfnamefont {C.}~\bibnamefont {Gong}}, \bibinfo {author} {\bibfnamefont {L.}~\bibnamefont {Lin}}, \bibinfo {author} {\bibfnamefont {L.}~\bibnamefont {Zhenglu}}, \bibinfo {author} {\bibfnamefont {J.}~\bibnamefont {Huiwen}}, \bibinfo {author} {\bibfnamefont {A.}~\bibnamefont {Stern}}, \bibinfo {author} {\bibfnamefont {X.}~\bibnamefont {Yang}}, \bibinfo {author} {\bibfnamefont {C.}~\bibnamefont {Ting}}, \bibinfo {author} {\bibfnamefont {B.}~\bibnamefont {Wei}}, \bibinfo {author} {\bibfnamefont {C.}~\bibnamefont {Wang}}, \bibinfo {author} {\bibfnamefont {Y.}~\bibnamefont {Wang}}, \bibinfo {author} {\bibfnamefont {Z.~Q.}\ \bibnamefont {Qui}}, \bibinfo {author} {\bibfnamefont {R.~J.}\ \bibnamefont {Cava}}, \bibinfo {author} {\bibfnamefont {S.~G.}\ \bibnamefont {Louie}}, \bibinfo {author} {\bibfnamefont {X.}~\bibnamefont {Jing}},\ and\ \bibinfo {author} {\bibfnamefont {Z.}~\bibnamefont {Xiang}},\ }\bibfield  {title} {\bibinfo {title} {{Discovery of intrinsic ferromagnetism
  in two-dimensional van der Waals crystals}},\ }\href@noop {} {\bibfield  {journal} {\bibinfo  {journal} {Nature}\ }\textbf {\bibinfo {volume} {546}},\ \bibinfo {pages} {265} (\bibinfo {year} {2017})}\BibitemShut {NoStop}%
\bibitem [{\citenamefont {Koyama}\ \emph {et~al.}(2000)\citenamefont {Koyama}, \citenamefont {Goto}, \citenamefont {Kanomata}, \citenamefont {Note},\ and\ \citenamefont {Takahashi}}]{Koyama2000FeSiImpurityMag}%
  \BibitemOpen
  \bibfield  {author} {\bibinfo {author} {\bibfnamefont {K.}~\bibnamefont {Koyama}}, \bibinfo {author} {\bibfnamefont {T.}~\bibnamefont {Goto}}, \bibinfo {author} {\bibfnamefont {T.}~\bibnamefont {Kanomata}}, \bibinfo {author} {\bibfnamefont {R.}~\bibnamefont {Note}},\ and\ \bibinfo {author} {\bibfnamefont {Y.}~\bibnamefont {Takahashi}},\ }\bibfield  {title} {\bibinfo {title} {{Nonlinear Magnetization Process of Single-Crystalline FeSi}},\ }\href@noop {} {\bibfield  {journal} {\bibinfo  {journal} {Journal of the Physical Society of Japan}\ }\textbf {\bibinfo {volume} {69}},\ \bibinfo {pages} {219} (\bibinfo {year} {2000})}\BibitemShut {NoStop}%
\bibitem [{\citenamefont {Renard}\ \emph {et~al.}(2005)\citenamefont {Renard}, \citenamefont {Gornyi}, \citenamefont {Tkachenko}, \citenamefont {Tkachenko}, \citenamefont {Kvon}, \citenamefont {Olshanetsky}, \citenamefont {Toropov},\ and\ \citenamefont {Portal}}]{Renard2005PhysRevB.72.075313dirty2DEG}%
  \BibitemOpen
  \bibfield  {author} {\bibinfo {author} {\bibfnamefont {V.~T.}\ \bibnamefont {Renard}}, \bibinfo {author} {\bibfnamefont {I.~V.}\ \bibnamefont {Gornyi}}, \bibinfo {author} {\bibfnamefont {O.~A.}\ \bibnamefont {Tkachenko}}, \bibinfo {author} {\bibfnamefont {V.~A.}\ \bibnamefont {Tkachenko}}, \bibinfo {author} {\bibfnamefont {Z.~D.}\ \bibnamefont {Kvon}}, \bibinfo {author} {\bibfnamefont {E.~B.}\ \bibnamefont {Olshanetsky}}, \bibinfo {author} {\bibfnamefont {A.~I.}\ \bibnamefont {Toropov}},\ and\ \bibinfo {author} {\bibfnamefont {J.-C.}\ \bibnamefont {Portal}},\ }\bibfield  {title} {\bibinfo {title} {{Quantum corrections to the conductivity and Hall coefficient of a two-dimensional electron gas in a dirty ${\mathrm{Al}}_{x}{\mathrm{Ga}}_{1\ensuremath{-}x}\mathrm{As}\mathrm{Ga}\mathrm{As}{\mathrm{Al}}_{x}{\mathrm{Ga}}_{1\ensuremath{-}x}\mathrm{As}$ quantum well: From the diffusive to the ballistic regime}},\ }\href {https://doi.org/10.1103/PhysRevB.72.075313} {\bibfield  {journal} {\bibinfo  {journal} {Phys.
  Rev. B}\ }\textbf {\bibinfo {volume} {72}},\ \bibinfo {pages} {075313} (\bibinfo {year} {2005})}\BibitemShut {NoStop}%
\bibitem [{\citenamefont {Chen}\ \emph {et~al.}(2011)\citenamefont {Chen}, \citenamefont {Li}, \citenamefont {Cullen}, \citenamefont {Williams},\ and\ \citenamefont {Fuhrer}}]{Chen2011NatPhysGrapheneKondo}%
  \BibitemOpen
  \bibfield  {author} {\bibinfo {author} {\bibfnamefont {J.-H.}\ \bibnamefont {Chen}}, \bibinfo {author} {\bibfnamefont {L.}~\bibnamefont {Li}}, \bibinfo {author} {\bibfnamefont {W.~G.}\ \bibnamefont {Cullen}}, \bibinfo {author} {\bibfnamefont {E.~D.}\ \bibnamefont {Williams}},\ and\ \bibinfo {author} {\bibfnamefont {M.~S.}\ \bibnamefont {Fuhrer}},\ }\bibfield  {title} {\bibinfo {title} {{Tunable Kondo effect in graphene with defects}},\ }\href@noop {} {\bibfield  {journal} {\bibinfo  {journal} {Nature Physics}\ }\textbf {\bibinfo {volume} {7}},\ \bibinfo {pages} {535} (\bibinfo {year} {2011})}\BibitemShut {NoStop}%
\bibitem [{\citenamefont {Nagaosa}\ \emph {et~al.}(2010)\citenamefont {Nagaosa}, \citenamefont {Sinova}, \citenamefont {Onoda}, \citenamefont {MacDonald},\ and\ \citenamefont {Ong}}]{Nagaosa2010RevModPhysAHEreview}%
  \BibitemOpen
  \bibfield  {author} {\bibinfo {author} {\bibfnamefont {N.}~\bibnamefont {Nagaosa}}, \bibinfo {author} {\bibfnamefont {J.}~\bibnamefont {Sinova}}, \bibinfo {author} {\bibfnamefont {S.}~\bibnamefont {Onoda}}, \bibinfo {author} {\bibfnamefont {A.~H.}\ \bibnamefont {MacDonald}},\ and\ \bibinfo {author} {\bibfnamefont {N.~P.}\ \bibnamefont {Ong}},\ }\bibfield  {title} {\bibinfo {title} {{Anomalous Hall effect}},\ }\href {https://doi.org/10.1103/RevModPhys.82.1539} {\bibfield  {journal} {\bibinfo  {journal} {Rev. Mod. Phys.}\ }\textbf {\bibinfo {volume} {82}},\ \bibinfo {pages} {1539} (\bibinfo {year} {2010})}\BibitemShut {NoStop}%
\bibitem [{\citenamefont {Kim}\ \emph {et~al.}(2013)\citenamefont {Kim}, \citenamefont {Thomas}, \citenamefont {Grant}, \citenamefont {Botimer}, \citenamefont {Fisk},\ and\ \citenamefont {Xia}}]{Kim2013SmB6WedgeHall}%
  \BibitemOpen
  \bibfield  {author} {\bibinfo {author} {\bibfnamefont {D.~J.}\ \bibnamefont {Kim}}, \bibinfo {author} {\bibfnamefont {S.}~\bibnamefont {Thomas}}, \bibinfo {author} {\bibfnamefont {T.}~\bibnamefont {Grant}}, \bibinfo {author} {\bibfnamefont {J.}~\bibnamefont {Botimer}}, \bibinfo {author} {\bibfnamefont {Z.}~\bibnamefont {Fisk}},\ and\ \bibinfo {author} {\bibfnamefont {J.}~\bibnamefont {Xia}},\ }\bibfield  {title} {\bibinfo {title} {{Surface Hall Effect and Nonlocal Transport in SmB$_6$: Evidence for Surface Conduction}},\ }\href@noop {} {\bibfield  {journal} {\bibinfo  {journal} {Sci. Rep.}\ }\textbf {\bibinfo {volume} {3}},\ \bibinfo {pages} {3150} (\bibinfo {year} {2013})}\BibitemShut {NoStop}%
\bibitem [{\citenamefont {Balasubramanian}\ \emph {et~al.}(2020)\citenamefont {Balasubramanian}, \citenamefont {Manchanda}, \citenamefont {Pahari}, \citenamefont {Chen}, \citenamefont {Zhang}, \citenamefont {Valloppilly}, \citenamefont {Li}, \citenamefont {Sarella}, \citenamefont {Yue}, \citenamefont {Ullah}, \citenamefont {Dev}, \citenamefont {Muller}, \citenamefont {Skomski}, \citenamefont {Hadjipanayis},\ and\ \citenamefont {Sellmyer}}]{Balas2020PhysRevLett.124.057201CoSiSkyrmions}%
  \BibitemOpen
  \bibfield  {author} {\bibinfo {author} {\bibfnamefont {B.}~\bibnamefont {Balasubramanian}}, \bibinfo {author} {\bibfnamefont {P.}~\bibnamefont {Manchanda}}, \bibinfo {author} {\bibfnamefont {R.}~\bibnamefont {Pahari}}, \bibinfo {author} {\bibfnamefont {Z.}~\bibnamefont {Chen}}, \bibinfo {author} {\bibfnamefont {W.}~\bibnamefont {Zhang}}, \bibinfo {author} {\bibfnamefont {S.~R.}\ \bibnamefont {Valloppilly}}, \bibinfo {author} {\bibfnamefont {X.}~\bibnamefont {Li}}, \bibinfo {author} {\bibfnamefont {A.}~\bibnamefont {Sarella}}, \bibinfo {author} {\bibfnamefont {L.}~\bibnamefont {Yue}}, \bibinfo {author} {\bibfnamefont {A.}~\bibnamefont {Ullah}}, \bibinfo {author} {\bibfnamefont {P.}~\bibnamefont {Dev}}, \bibinfo {author} {\bibfnamefont {D.~A.}\ \bibnamefont {Muller}}, \bibinfo {author} {\bibfnamefont {R.}~\bibnamefont {Skomski}}, \bibinfo {author} {\bibfnamefont {G.~C.}\ \bibnamefont {Hadjipanayis}},\ and\ \bibinfo {author} {\bibfnamefont {D.~J.}\ \bibnamefont {Sellmyer}},\ }\bibfield  {title} {\bibinfo
  {title} {{Chiral Magnetism and High-Temperature Skyrmions in B20-Ordered Co-Si}},\ }\href {https://doi.org/10.1103/PhysRevLett.124.057201} {\bibfield  {journal} {\bibinfo  {journal} {Phys. Rev. Lett.}\ }\textbf {\bibinfo {volume} {124}},\ \bibinfo {pages} {057201} (\bibinfo {year} {2020})}\BibitemShut {NoStop}%
\bibitem [{\citenamefont {M\"unzer}\ \emph {et~al.}(2010)\citenamefont {M\"unzer}, \citenamefont {Neubauer}, \citenamefont {Adams}, \citenamefont {M\"uhlbauer}, \citenamefont {Franz}, \citenamefont {Jonietz}, \citenamefont {Georgii}, \citenamefont {B\"oni}, \citenamefont {Pedersen}, \citenamefont {Schmidt}, \citenamefont {Rosch},\ and\ \citenamefont {Pfleiderer}}]{Munzer2010PhysRevB.81.041203FeCoSiSkyrmion}%
  \BibitemOpen
  \bibfield  {author} {\bibinfo {author} {\bibfnamefont {W.}~\bibnamefont {M\"unzer}}, \bibinfo {author} {\bibfnamefont {A.}~\bibnamefont {Neubauer}}, \bibinfo {author} {\bibfnamefont {T.}~\bibnamefont {Adams}}, \bibinfo {author} {\bibfnamefont {S.}~\bibnamefont {M\"uhlbauer}}, \bibinfo {author} {\bibfnamefont {C.}~\bibnamefont {Franz}}, \bibinfo {author} {\bibfnamefont {F.}~\bibnamefont {Jonietz}}, \bibinfo {author} {\bibfnamefont {R.}~\bibnamefont {Georgii}}, \bibinfo {author} {\bibfnamefont {P.}~\bibnamefont {B\"oni}}, \bibinfo {author} {\bibfnamefont {B.}~\bibnamefont {Pedersen}}, \bibinfo {author} {\bibfnamefont {M.}~\bibnamefont {Schmidt}}, \bibinfo {author} {\bibfnamefont {A.}~\bibnamefont {Rosch}},\ and\ \bibinfo {author} {\bibfnamefont {C.}~\bibnamefont {Pfleiderer}},\ }\bibfield  {title} {\bibinfo {title} {{Skyrmion lattice in the doped semiconductor ${\text{Fe}}_{1\ensuremath{-}x}{\text{Co}}_{x}\text{Si}$}},\ }\href {https://doi.org/10.1103/PhysRevB.81.041203} {\bibfield  {journal} {\bibinfo
  {journal} {Phys. Rev. B}\ }\textbf {\bibinfo {volume} {81}},\ \bibinfo {pages} {041203} (\bibinfo {year} {2010})}\BibitemShut {NoStop}%
\bibitem [{\citenamefont {Bauer}\ \emph {et~al.}(2016)\citenamefont {Bauer}, \citenamefont {Garst},\ and\ \citenamefont {Pfleiderer}}]{Bauer2016PhysRevB.93.235144FeCoSiMagHistory}%
  \BibitemOpen
  \bibfield  {author} {\bibinfo {author} {\bibfnamefont {A.}~\bibnamefont {Bauer}}, \bibinfo {author} {\bibfnamefont {M.}~\bibnamefont {Garst}},\ and\ \bibinfo {author} {\bibfnamefont {C.}~\bibnamefont {Pfleiderer}},\ }\bibfield  {title} {\bibinfo {title} {{History dependence of the magnetic properties of single-crystal ${\mathrm{Fe}}_{1\ensuremath{-}x}{\mathrm{Co}}_{x}\mathrm{Si}$}},\ }\href {https://doi.org/10.1103/PhysRevB.93.235144} {\bibfield  {journal} {\bibinfo  {journal} {Phys. Rev. B}\ }\textbf {\bibinfo {volume} {93}},\ \bibinfo {pages} {235144} (\bibinfo {year} {2016})}\BibitemShut {NoStop}%
\bibitem [{\citenamefont {Manyala}\ \emph {et~al.}(2004)\citenamefont {Manyala}, \citenamefont {Sidis}, \citenamefont {DiTusa}, \citenamefont {Aeppli}, \citenamefont {Young},\ and\ \citenamefont {Fisk}}]{Manyala2004NatMatFeSiAHE}%
  \BibitemOpen
  \bibfield  {author} {\bibinfo {author} {\bibfnamefont {N.}~\bibnamefont {Manyala}}, \bibinfo {author} {\bibfnamefont {Y.}~\bibnamefont {Sidis}}, \bibinfo {author} {\bibfnamefont {J.~F.}\ \bibnamefont {DiTusa}}, \bibinfo {author} {\bibfnamefont {G.}~\bibnamefont {Aeppli}}, \bibinfo {author} {\bibfnamefont {D.~P.}\ \bibnamefont {Young}},\ and\ \bibinfo {author} {\bibfnamefont {Z.}~\bibnamefont {Fisk}},\ }\bibfield  {title} {\bibinfo {title} {{Large anomalous Hall effect in a silicon-based magnetic semiconductor}},\ }\href@noop {} {\bibfield  {journal} {\bibinfo  {journal} {Nature Materials}\ }\textbf {\bibinfo {volume} {3}},\ \bibinfo {pages} {255} (\bibinfo {year} {2004})}\BibitemShut {NoStop}%
\bibitem [{\citenamefont {Beille}\ \emph {et~al.}(1981)\citenamefont {Beille}, \citenamefont {Voiron}, \citenamefont {Towfiq}, \citenamefont {Roth},\ and\ \citenamefont {Zhang}}]{Beille_1981_JPhysF_FeCoSiHelical}%
  \BibitemOpen
  \bibfield  {author} {\bibinfo {author} {\bibfnamefont {J.}~\bibnamefont {Beille}}, \bibinfo {author} {\bibfnamefont {J.}~\bibnamefont {Voiron}}, \bibinfo {author} {\bibfnamefont {F.}~\bibnamefont {Towfiq}}, \bibinfo {author} {\bibfnamefont {M.}~\bibnamefont {Roth}},\ and\ \bibinfo {author} {\bibfnamefont {Z.~Y.}\ \bibnamefont {Zhang}},\ }\bibfield  {title} {\bibinfo {title} {{Helimagnetic structure of the Fe$_x$Co$_{1-x}$Si alloys}},\ }\href {https://doi.org/10.1088/0305-4608/11/10/026} {\bibfield  {journal} {\bibinfo  {journal} {Journal of Physics F: Metal Physics}\ }\textbf {\bibinfo {volume} {11}},\ \bibinfo {pages} {2153} (\bibinfo {year} {1981})}\BibitemShut {NoStop}%
\bibitem [{\citenamefont {Takeda}\ \emph {et~al.}(2009)\citenamefont {Takeda}, \citenamefont {Endoh}, \citenamefont {Kakurai}, \citenamefont {Onose}, \citenamefont {Suzuki},\ and\ \citenamefont {Tokura}}]{Takeda_2009_JPSJ_FeCoSi_SANS}%
  \BibitemOpen
  \bibfield  {author} {\bibinfo {author} {\bibfnamefont {M.}~\bibnamefont {Takeda}}, \bibinfo {author} {\bibfnamefont {Y.}~\bibnamefont {Endoh}}, \bibinfo {author} {\bibfnamefont {K.}~\bibnamefont {Kakurai}}, \bibinfo {author} {\bibfnamefont {Y.}~\bibnamefont {Onose}}, \bibinfo {author} {\bibfnamefont {J.}~\bibnamefont {Suzuki}},\ and\ \bibinfo {author} {\bibfnamefont {Y.}~\bibnamefont {Tokura}},\ }\bibfield  {title} {\bibinfo {title} {{Nematic-to-Smectic Transition of Magnetic Texture in Conical State}},\ }\href {https://doi.org/10.1143/JPSJ.78.093704} {\bibfield  {journal} {\bibinfo  {journal} {Journal of the Physical Society of Japan}\ }\textbf {\bibinfo {volume} {78}},\ \bibinfo {pages} {093704} (\bibinfo {year} {2009})},\ \Eprint {https://arxiv.org/abs/https://doi.org/10.1143/JPSJ.78.093704} {https://doi.org/10.1143/JPSJ.78.093704} \BibitemShut {NoStop}%
\bibitem [{\citenamefont {Prakash}\ \emph {et~al.}(2007)\citenamefont {Prakash}, \citenamefont {Choudhary}, \citenamefont {Chandra}, \citenamefont {Lakshmi},\ and\ \citenamefont {Phase}}]{Prakash2007JPhysCondMat}%
  \BibitemOpen
  \bibfield  {author} {\bibinfo {author} {\bibfnamefont {R.}~\bibnamefont {Prakash}}, \bibinfo {author} {\bibfnamefont {R.~J.}\ \bibnamefont {Choudhary}}, \bibinfo {author} {\bibfnamefont {L.~S.~S.}\ \bibnamefont {Chandra}}, \bibinfo {author} {\bibfnamefont {N.}~\bibnamefont {Lakshmi}},\ and\ \bibinfo {author} {\bibfnamefont {D.~M.}\ \bibnamefont {Phase}},\ }\bibfield  {title} {\bibinfo {title} {{Electrical and magnetic transport properties of Fe$_3$O$_4$ thin films on a GaAs(100) substrate}},\ }\href {https://doi.org/10.1088/0953-8984/19/48/486212} {\bibfield  {journal} {\bibinfo  {journal} {Journal of Physics: Condensed Matter}\ }\textbf {\bibinfo {volume} {19}},\ \bibinfo {pages} {486212} (\bibinfo {year} {2007})}\BibitemShut {NoStop}%
\bibitem [{\citenamefont {Hori}\ \emph {et~al.}(2023)\citenamefont {Hori}, \citenamefont {Kanazawa}, \citenamefont {Hirayama}, \citenamefont {Fujiwara}, \citenamefont {Tsukazaki}, \citenamefont {Ichikawa}, \citenamefont {Kawasaki},\ and\ \citenamefont {Tokura}}]{Hori_2023_AdvMatFeSiCapping}%
  \BibitemOpen
  \bibfield  {author} {\bibinfo {author} {\bibfnamefont {T.}~\bibnamefont {Hori}}, \bibinfo {author} {\bibfnamefont {N.}~\bibnamefont {Kanazawa}}, \bibinfo {author} {\bibfnamefont {M.}~\bibnamefont {Hirayama}}, \bibinfo {author} {\bibfnamefont {K.}~\bibnamefont {Fujiwara}}, \bibinfo {author} {\bibfnamefont {A.}~\bibnamefont {Tsukazaki}}, \bibinfo {author} {\bibfnamefont {M.}~\bibnamefont {Ichikawa}}, \bibinfo {author} {\bibfnamefont {M.}~\bibnamefont {Kawasaki}},\ and\ \bibinfo {author} {\bibfnamefont {Y.}~\bibnamefont {Tokura}},\ }\bibfield  {title} {\bibinfo {title} {{A Noble-Metal-Free Spintronic System with Proximity-Enhanced Ferromagnetic Topological Surface State of FeSi above Room Temperature}},\ }\href {https://doi.org/https://doi.org/10.1002/adma.202206801} {\bibfield  {journal} {\bibinfo  {journal} {Advanced Materials}\ }\textbf {\bibinfo {volume} {35}},\ \bibinfo {pages} {2206801} (\bibinfo {year} {2023})},\ \Eprint
  {https://arxiv.org/abs/https://onlinelibrary.wiley.com/doi/pdf/10.1002/adma.202206801} {https://onlinelibrary.wiley.com/doi/pdf/10.1002/adma.202206801} \BibitemShut {NoStop}%
\bibitem [{\citenamefont {Hori}\ \emph {et~al.}(2024)\citenamefont {Hori}, \citenamefont {Kanazawa}, \citenamefont {Matsuura}, \citenamefont {Ishizuka}, \citenamefont {Fujiwara}, \citenamefont {Tsukazaki}, \citenamefont {Ichikawa}, \citenamefont {Kawasaki}, \citenamefont {Kagawa}, \citenamefont {Hirayama},\ and\ \citenamefont {Tokura}}]{Hori2024PRM_PtCapFeSi}%
  \BibitemOpen
  \bibfield  {author} {\bibinfo {author} {\bibfnamefont {T.}~\bibnamefont {Hori}}, \bibinfo {author} {\bibfnamefont {N.}~\bibnamefont {Kanazawa}}, \bibinfo {author} {\bibfnamefont {K.}~\bibnamefont {Matsuura}}, \bibinfo {author} {\bibfnamefont {H.}~\bibnamefont {Ishizuka}}, \bibinfo {author} {\bibfnamefont {K.}~\bibnamefont {Fujiwara}}, \bibinfo {author} {\bibfnamefont {A.}~\bibnamefont {Tsukazaki}}, \bibinfo {author} {\bibfnamefont {M.}~\bibnamefont {Ichikawa}}, \bibinfo {author} {\bibfnamefont {M.}~\bibnamefont {Kawasaki}}, \bibinfo {author} {\bibfnamefont {F.}~\bibnamefont {Kagawa}}, \bibinfo {author} {\bibfnamefont {M.}~\bibnamefont {Hirayama}},\ and\ \bibinfo {author} {\bibfnamefont {Y.}~\bibnamefont {Tokura}},\ }\bibfield  {title} {\bibinfo {title} {{Strongly pinned skyrmionic bubbles and higher-order nonlinear Hall resistances at the interface of Pt/FeSi bilayer}},\ }\href {https://doi.org/10.1103/PhysRevMaterials.8.044407} {\bibfield  {journal} {\bibinfo  {journal} {Phys. Rev. Mater.}\ }\textbf
  {\bibinfo {volume} {8}},\ \bibinfo {pages} {044407} (\bibinfo {year} {2024})}\BibitemShut {NoStop}%
\bibitem [{\citenamefont {Manyala}\ \emph {et~al.}(2009)\citenamefont {Manyala}, \citenamefont {Ngom}, \citenamefont {Beye}, \citenamefont {Bucher}, \citenamefont {Maaza}, \citenamefont {Strydom}, \citenamefont {Forbes}, \citenamefont {Johnson},\ and\ \citenamefont {DiTusa}}]{Manyala2009APLThinFilmFeSi10.1063/1.3152766}%
  \BibitemOpen
  \bibfield  {author} {\bibinfo {author} {\bibfnamefont {N.}~\bibnamefont {Manyala}}, \bibinfo {author} {\bibfnamefont {B.~D.}\ \bibnamefont {Ngom}}, \bibinfo {author} {\bibfnamefont {A.~C.}\ \bibnamefont {Beye}}, \bibinfo {author} {\bibfnamefont {R.}~\bibnamefont {Bucher}}, \bibinfo {author} {\bibfnamefont {M.}~\bibnamefont {Maaza}}, \bibinfo {author} {\bibfnamefont {A.}~\bibnamefont {Strydom}}, \bibinfo {author} {\bibfnamefont {A.}~\bibnamefont {Forbes}}, \bibinfo {author} {\bibfnamefont {J.}~\bibnamefont {Johnson}, \bibfnamefont {A.~T.~Charlie}},\ and\ \bibinfo {author} {\bibfnamefont {J.~F.}\ \bibnamefont {DiTusa}},\ }\bibfield  {title} {\bibinfo {title} {{Structural and magnetic properties of $\epsilon$-Fe$_{1-x}$Co$_x$Si thin films deposited via pulsed laser deposition}},\ }\href {https://doi.org/10.1063/1.3152766} {\bibfield  {journal} {\bibinfo  {journal} {Applied Physics Letters}\ }\textbf {\bibinfo {volume} {94}},\ \bibinfo {pages} {232503} (\bibinfo {year} {2009})}\BibitemShut {NoStop}%
\bibitem [{\citenamefont {Breindel}\ \emph {et~al.}(2023)\citenamefont {Breindel}, \citenamefont {Deng}, \citenamefont {Moir}, \citenamefont {Fang}, \citenamefont {Ran}, \citenamefont {Lou}, \citenamefont {Li}, \citenamefont {Zeng}, \citenamefont {Shu}, \citenamefont {Wolowiec}, \citenamefont {Schuller}, \citenamefont {Rosa}, \citenamefont {Fisk}, \citenamefont {Singleton},\ and\ \citenamefont {Maple}}]{Briendel2023ProbFeSiPublish}%
  \BibitemOpen
  \bibfield  {author} {\bibinfo {author} {\bibfnamefont {A.}~\bibnamefont {Breindel}}, \bibinfo {author} {\bibfnamefont {Y.}~\bibnamefont {Deng}}, \bibinfo {author} {\bibfnamefont {C.~M.}\ \bibnamefont {Moir}}, \bibinfo {author} {\bibfnamefont {Y.}~\bibnamefont {Fang}}, \bibinfo {author} {\bibfnamefont {S.}~\bibnamefont {Ran}}, \bibinfo {author} {\bibfnamefont {H.}~\bibnamefont {Lou}}, \bibinfo {author} {\bibfnamefont {S.}~\bibnamefont {Li}}, \bibinfo {author} {\bibfnamefont {Q.}~\bibnamefont {Zeng}}, \bibinfo {author} {\bibfnamefont {L.}~\bibnamefont {Shu}}, \bibinfo {author} {\bibfnamefont {C.~T.}\ \bibnamefont {Wolowiec}}, \bibinfo {author} {\bibfnamefont {I.~K.}\ \bibnamefont {Schuller}}, \bibinfo {author} {\bibfnamefont {P.~F.~S.}\ \bibnamefont {Rosa}}, \bibinfo {author} {\bibfnamefont {Z.}~\bibnamefont {Fisk}}, \bibinfo {author} {\bibfnamefont {J.}~\bibnamefont {Singleton}},\ and\ \bibinfo {author} {\bibfnamefont {M.~B.}\ \bibnamefont {Maple}},\ }\bibfield  {title} {\bibinfo {title} {{Probing FeSi, a
  d-electron topological Kondo insulator candidate, with magnetic field, pressure, and microwaves}},\ }\href@noop {} {\bibfield  {journal} {\bibinfo  {journal} {Proceedings of the National Academy of Sciences}\ }\textbf {\bibinfo {volume} {120}},\ \bibinfo {pages} {e2216367120} (\bibinfo {year} {2023})}\BibitemShut {NoStop}%
\bibitem [{\citenamefont {Deng}\ \emph {et~al.}(2023)\citenamefont {Deng}, \citenamefont {Yan}, \citenamefont {Wang}, \citenamefont {Lee-Wong}, \citenamefont {Moir}, \citenamefont {Fang}, \citenamefont {Xie},\ and\ \citenamefont {Maple}}]{Deng2023PhysRevB.108.115158AngleResMR_Sn_FeSi}%
  \BibitemOpen
  \bibfield  {author} {\bibinfo {author} {\bibfnamefont {Y.}~\bibnamefont {Deng}}, \bibinfo {author} {\bibfnamefont {Y.}~\bibnamefont {Yan}}, \bibinfo {author} {\bibfnamefont {H.}~\bibnamefont {Wang}}, \bibinfo {author} {\bibfnamefont {E.}~\bibnamefont {Lee-Wong}}, \bibinfo {author} {\bibfnamefont {C.~M.}\ \bibnamefont {Moir}}, \bibinfo {author} {\bibfnamefont {Y.}~\bibnamefont {Fang}}, \bibinfo {author} {\bibfnamefont {W.}~\bibnamefont {Xie}},\ and\ \bibinfo {author} {\bibfnamefont {M.~B.}\ \bibnamefont {Maple}},\ }\bibfield  {title} {\bibinfo {title} {{Possible surface magnetism in the topological Kondo insulator candidate FeSi}},\ }\href {https://doi.org/10.1103/PhysRevB.108.115158} {\bibfield  {journal} {\bibinfo  {journal} {Phys. Rev. B}\ }\textbf {\bibinfo {volume} {108}},\ \bibinfo {pages} {115158} (\bibinfo {year} {2023})}\BibitemShut {NoStop}%
\bibitem [{\citenamefont {Brando}\ \emph {et~al.}(2016)\citenamefont {Brando}, \citenamefont {Belitz}, \citenamefont {Grosche},\ and\ \citenamefont {Kirkpatrick}}]{Brando2016RevModPhys.88.025006QCP}%
  \BibitemOpen
  \bibfield  {author} {\bibinfo {author} {\bibfnamefont {M.}~\bibnamefont {Brando}}, \bibinfo {author} {\bibfnamefont {D.}~\bibnamefont {Belitz}}, \bibinfo {author} {\bibfnamefont {F.~M.}\ \bibnamefont {Grosche}},\ and\ \bibinfo {author} {\bibfnamefont {T.~R.}\ \bibnamefont {Kirkpatrick}},\ }\bibfield  {title} {\bibinfo {title} {{Metallic quantum ferromagnets}},\ }\href {https://doi.org/10.1103/RevModPhys.88.025006} {\bibfield  {journal} {\bibinfo  {journal} {Rev. Mod. Phys.}\ }\textbf {\bibinfo {volume} {88}},\ \bibinfo {pages} {025006} (\bibinfo {year} {2016})}\BibitemShut {NoStop}%
\bibitem [{\citenamefont {Saha}\ \emph {et~al.}(2022)\citenamefont {Saha}, \citenamefont {Dutta}, \citenamefont {Gupta}, \citenamefont {Bandyopadhyay},\ and\ \citenamefont {Das}}]{Saha2022PhysRevB.105.214407GrittithsManganite}%
  \BibitemOpen
  \bibfield  {author} {\bibinfo {author} {\bibfnamefont {S.}~\bibnamefont {Saha}}, \bibinfo {author} {\bibfnamefont {A.}~\bibnamefont {Dutta}}, \bibinfo {author} {\bibfnamefont {S.}~\bibnamefont {Gupta}}, \bibinfo {author} {\bibfnamefont {S.}~\bibnamefont {Bandyopadhyay}},\ and\ \bibinfo {author} {\bibfnamefont {I.}~\bibnamefont {Das}},\ }\bibfield  {title} {\bibinfo {title} {{Origin of the Griffiths phase and correlation with the magnetic phase transition in the nanocrystalline manganite ${\mathrm{La}}_{0.4}{({\mathrm{Ca}}_{0.5}{\mathrm{Sr}}_{0.5})}_{0.6}{\mathrm{MnO}}_{3}$}},\ }\href {https://doi.org/10.1103/PhysRevB.105.214407} {\bibfield  {journal} {\bibinfo  {journal} {Phys. Rev. B}\ }\textbf {\bibinfo {volume} {105}},\ \bibinfo {pages} {214407} (\bibinfo {year} {2022})}\BibitemShut {NoStop}%
\bibitem [{\citenamefont {Das}\ \emph {et~al.}(2018)\citenamefont {Das}, \citenamefont {Banu}, \citenamefont {Das},\ and\ \citenamefont {Dev}}]{DAS2018SolStateComm36SmMnO3DoppedGriffiths}%
  \BibitemOpen
  \bibfield  {author} {\bibinfo {author} {\bibfnamefont {K.}~\bibnamefont {Das}}, \bibinfo {author} {\bibfnamefont {N.}~\bibnamefont {Banu}}, \bibinfo {author} {\bibfnamefont {I.}~\bibnamefont {Das}},\ and\ \bibinfo {author} {\bibfnamefont {B.}~\bibnamefont {Dev}},\ }\bibfield  {title} {\bibinfo {title} {{Magnetocaloric effect of polycrystalline Sm0.5Ca0.5MnO3 compound: Investigation of low temperature magnetic state}},\ }\href {https://doi.org/https://doi.org/10.1016/j.ssc.2018.02.014} {\bibfield  {journal} {\bibinfo  {journal} {Solid State Communications}\ }\textbf {\bibinfo {volume} {274}},\ \bibinfo {pages} {36} (\bibinfo {year} {2018})}\BibitemShut {NoStop}%
\bibitem [{\citenamefont {Ohnuma}\ \emph {et~al.}(2012)\citenamefont {Ohnuma}, \citenamefont {Abe}, \citenamefont {Shimenouchi}, \citenamefont {Omori}, \citenamefont {Kainuma},\ and\ \citenamefont {Ishida}}]{Ohnuma2012ISIJFeSiMetallurgy}%
  \BibitemOpen
  \bibfield  {author} {\bibinfo {author} {\bibfnamefont {I.}~\bibnamefont {Ohnuma}}, \bibinfo {author} {\bibfnamefont {S.}~\bibnamefont {Abe}}, \bibinfo {author} {\bibfnamefont {S.}~\bibnamefont {Shimenouchi}}, \bibinfo {author} {\bibfnamefont {T.}~\bibnamefont {Omori}}, \bibinfo {author} {\bibfnamefont {R.}~\bibnamefont {Kainuma}},\ and\ \bibinfo {author} {\bibfnamefont {K.}~\bibnamefont {Ishida}},\ }\bibfield  {title} {\bibinfo {title} {{Experimental and Thermodynamic Studies of the Fe and Si Binary System}},\ }\href {https://doi.org/10.2355/isijinternational.52.540} {\bibfield  {journal} {\bibinfo  {journal} {ISIJ International}\ }\textbf {\bibinfo {volume} {52}},\ \bibinfo {pages} {540} (\bibinfo {year} {2012})}\BibitemShut {NoStop}%
\bibitem [{\citenamefont {Alexandrov}\ \emph {et~al.}(2015)\citenamefont {Alexandrov}, \citenamefont {Coleman},\ and\ \citenamefont {Erten}}]{Alexandrov2015PhysRevLett.114.177202KondoBreakdown}%
  \BibitemOpen
  \bibfield  {author} {\bibinfo {author} {\bibfnamefont {V.}~\bibnamefont {Alexandrov}}, \bibinfo {author} {\bibfnamefont {P.}~\bibnamefont {Coleman}},\ and\ \bibinfo {author} {\bibfnamefont {O.}~\bibnamefont {Erten}},\ }\bibfield  {title} {\bibinfo {title} {{Kondo Breakdown in Topological Kondo Insulators}},\ }\href {https://doi.org/10.1103/PhysRevLett.114.177202} {\bibfield  {journal} {\bibinfo  {journal} {Phys. Rev. Lett.}\ }\textbf {\bibinfo {volume} {114}},\ \bibinfo {pages} {177202} (\bibinfo {year} {2015})}\BibitemShut {NoStop}%
\bibitem [{\citenamefont {Nakajima}\ \emph {et~al.}(2016)\citenamefont {Nakajima}, \citenamefont {Syers}, \citenamefont {Wang}, \citenamefont {Wang},\ and\ \citenamefont {Paglione}}]{Nakajima2016NatPhysSmB6Ferromagentism}%
  \BibitemOpen
  \bibfield  {author} {\bibinfo {author} {\bibfnamefont {Y.}~\bibnamefont {Nakajima}}, \bibinfo {author} {\bibfnamefont {P.}~\bibnamefont {Syers}}, \bibinfo {author} {\bibfnamefont {X.}~\bibnamefont {Wang}}, \bibinfo {author} {\bibfnamefont {R.}~\bibnamefont {Wang}},\ and\ \bibinfo {author} {\bibfnamefont {J.}~\bibnamefont {Paglione}},\ }\bibfield  {title} {\bibinfo {title} {{One-dimensional edge state transport in a topological Kondo insulator}},\ }\href@noop {} {\bibfield  {journal} {\bibinfo  {journal} {Nature Physics}\ }\textbf {\bibinfo {volume} {12}},\ \bibinfo {pages} {213} (\bibinfo {year} {2016})}\BibitemShut {NoStop}%
\bibitem [{\citenamefont {Sun}\ \emph {et~al.}(2017)\citenamefont {Sun}, \citenamefont {Zhuo}, \citenamefont {Wu},\ and\ \citenamefont {Yang}}]{Sun2017Fe2SiThinFilm}%
  \BibitemOpen
  \bibfield  {author} {\bibinfo {author} {\bibfnamefont {Y.}~\bibnamefont {Sun}}, \bibinfo {author} {\bibfnamefont {Z.}~\bibnamefont {Zhuo}}, \bibinfo {author} {\bibfnamefont {X.}~\bibnamefont {Wu}},\ and\ \bibinfo {author} {\bibfnamefont {J.}~\bibnamefont {Yang}},\ }\bibfield  {title} {\bibinfo {title} {{Room-Temperature Ferromagnetism in Two-Dimensional Fe$_2$Si Nanosheet with Enhanced Spin-Polarization Ratio}},\ }\href {https://doi.org/10.1021/acs.nanolett.6b04884} {\bibfield  {journal} {\bibinfo  {journal} {Nano Letters}\ }\textbf {\bibinfo {volume} {17}},\ \bibinfo {pages} {2771} (\bibinfo {year} {2017})},\ \bibinfo {note} {pMID: 28441496},\ \Eprint {https://arxiv.org/abs/https://doi.org/10.1021/acs.nanolett.6b04884} {https://doi.org/10.1021/acs.nanolett.6b04884} \BibitemShut {NoStop}%
\end{thebibliography}
%

\end{document}